\documentclass[12pt,compress]{elsarticle}
\usepackage{amsmath,amssymb,amscd}
\usepackage{enumitem}
\usepackage{float}
\usepackage{blindtext}

\usepackage{color}
\usepackage{mathrsfs}
\usepackage{mathabx}
\usepackage{ytableau}
\usepackage{hyperref}
\usepackage{tikz}
\usetikzlibrary{shapes.callouts}

\journal{Nuclear Physics B}

\newcommand{\beq}{\begin{equation}}
\newcommand{\eeq}{\end{equation}}
\newcommand{\bea}{\begin{eqnarray}}
\newcommand{\eea}{\end{eqnarray}}
\newcommand{\bse}{\begin{subequations}}
\newcommand{\ese}{\end{subequations}}

\newcommand{\noi}{\noindent}

\newcommand{\nn}{\nonumber}

\newcommand{\ba}{\begin{array}}
\newcommand{\ea}{\end{array}}
\newcommand{\balign}{\begin{align}}
\newcommand{\ealign}{\end{align}}

\newcommand{\al}{\alpha}

\newcommand{\de}{\delta}
\newcommand{\ep}{\epsilon}

\newcommand{\mbf}[1]{\mathbf{#1}}

\newcommand{\mbs}[1]{\boldsymbol{#1}}

\newcommand{\mbb}[1]{\mathbb{#1}}

\newcommand{\mc}[1]{\mathcal{#1}}

\newcommand{\msc}[1]{\mathscr{#1}}

\newcommand{\mr}[1]{\mathrm{#1}}

\newcommand{\wt}[1]{\widetilde{#1}}

\newcommand{\one}{1\hskip -1mm{\rm l}}

\renewcommand{\leq}{\leqslant}
\renewcommand{\geq}{\geqslant}

\renewcommand{\v}{\vert}
\newcommand{\half}{$\frac{1}{2}~$}
\newcommand{\su}[1]{su($#1$)}

\newcommand\ket[1]{|#1\rangle}

\newcommand{\Z}{\mbb{Z}_{\geq 0}}
\newcommand{\eq}[1]{(\ref{#1})}
\newcommand{\Eq}[1]{Eq.~(\ref{#1})}
\begin{document}
% \begin{flushright} hep-th/9701097
% \end{flushright}
% \rightline{TIFR/TH/97-01}
% \rightline{January 1997}

\begin{center}
{\large \bf \sf %Analysis of  the 
Appearance of branched motifs in the spectra of  \\{$BC_N$} type  Polychronakos  spin chains }
%through Rogers-Szeg\"o polynomials }

\vspace{1.3cm}
{\sf 
Bireswar Basu-Mallick$^1$\footnote{
Corresponding author, Fax:+91-33-2337-4637, 
Telephone:+91-33-2337-5345, 
E-mail address: bireswar.basumallick@saha.ac.in}
and Madhurima Sinha$^1$\footnote
{E-mail~address: madhurima.sinha@saha.ac.in},
}

\bigskip

{\em $^1$Theory Division, Saha Institute of Nuclear Physics, HBNI,\\
1/AF Bidhan Nagar, Kolkata 700 064, India}

\end{center}

\bigskip \bigskip

\noi {\bf Abstract:} 
\vspace {.2 cm}

As is well known, energy levels appearing in the highly degenerate spectra of the $A_{N-1}$ type of  
Haldane-Shastry and Polychronakos spin chains can be classified through the motifs,  which are characterized by some sequences of the binary digits like `0' and `1'. In a similar way, at present we  classify all energy levels appearing in the spectra of the $BC_N$ type of Polychronakos spin chains with Hamiltonians containing supersymmetric analogue of polarized spin reversal operators. To this end, we show that the $BC_N$  type of multivariate super Rogers-Szeg\"o (SRS) polynomials, which at a certain limit reduce to the partition functions of the later type of Polychronakos  spin chains, satisfy some recursion relation involving a $q$-deformation of the  elementary supersymmetric  polynomials.
Subsequently, we use a Jacobi-Trudi like formula to define the corresponding $q$-deformed super Schur polynomials and derive a novel expression for the $BC_N$ type of multivariate SRS polynomials as suitable linear combinations of the $q$-deformed super Schur polynomials. Such an expression for SRS polynomials leads to a complete classification of all energy levels appearing in the spectra of the $BC_N$ type of Polychronakos spin chains through the  `branched' motifs, which are characterized by some sequences of integers of the form $(\de_1, \de_2,...,\de_{N-1}|l)$, where $\de_i \in \{ 0,1 \}$ and $ l \in \{ 0,1,...,N \}$. Finally, we 
derive an extended boson-fermion duality relation among the restricted super Schur polynomials and show that the partition functions of  the $BC_N$ type of Polychronakos spin chains also exhibit similar type of duality relation.

\vspace{.74 cm}
\noi PACS No.: 02.30.Ik, 05.30.-d, 75.10.Pq, 75.10.Jm 
%02.30.Ik, 03.65.Ge, 02.10.De, 11.10.Lm

\vspace {.2 cm}

\vspace {.65 cm}
\begin{center}
{\it Dedicated to Artemio Gonz\'alez-L\'opez on the occasion of 
his 60th birthday}
\end{center}

%Coordinate Bethe ansatz; Derivative delta-function Bose 
%gas; Clusters of bound particles; Farey sequence 

\newpage

\baselineskip 16 pt

\noi \section{Introduction}
\renewcommand{\theequation}{1.{\arabic{equation}}}
\setcounter{equation}{0}
\medskip
 The interplay between symmetry and exact solvability in one-dimensional quantum 
 spin chains with long-range interactions 
%As is well known, the study of exactly solvable and quantum integrable spin chains with long-range 
%interactions in one dimension~ %and their supersymmetric extensions~
\cite{Ha88,
Sh88,HHTBP92,BGHP93,Ha93, SS93,
Po93,Fr93,Po94,Hi95npb,KKN97,BUW99,HB00,FG05,BB06,BBHS07,BBS08, BBH10,EFG12,FGLR18,
BPS95,YT96,
EFG10, EFGR05,BFGR08,BFGR09,KK09} 
%has connected apparently diverse topics  
has recently attracted a lot of attention due to its relevance  
in apparently diverse topics
of  physics and mathematics. 
Indeed, there are many important applications of this type of spin chains in various subjects like quantum electric transport phenomena~\cite{BR94,Ca95},
condensed matter systems obeying generalized exclusion
statistics~\cite{Ha93,KK09,Po06},
random matrix theory~\cite{TSA95}, 
`infinite matrix product states' in conformal field theory~\cite{CS10,NCS11,TNS14,BQ14,TS15,BFG16},
planar ${\mc N}=4$ super Yang--Mills theory~\cite{BKS03,BBL09,Bea12} and 
%quantum Hall effect~\cite{AI94}, 
Yangian quantum groups~\cite{BGHP93,Ha93,Hi95npb,BBHS07, BBH10}.
%and Rogers-Szeg\"o (RS) polynomials~\cite{Hi95npb,Hi95,Hi97,KKN97,HB00}. 
The study of such exactly solvable spin chains with long-range interactions 
was initiated through the pioneering works of 
Haldane and Shastry~\cite{Ha88,Sh88}. 
They derived the exact spectrum of a spin-\half chain with lattice sites equally spaced on a circle and spins interacting with each other through a pairwise exchange interaction, inversely proportional to the square of their chord distances.
An astonishing feature of this \su{2} Haldane-Shastry (HS) 
spin chain and its \su{m|n} supersymmetric generalizations is that they exhibit Yangian quantum group symmetry even for finite number of lattice sites. 
As a result, the degenerate multiplets (formed due to Yangian symmetry)  
in the energy spectra of these spin chains can be classified in an efficient way by using specific sequences of the binary digits `0' and `1', which 
are known as `motifs' in the literature~\cite{HHTBP92,BGHP93}.
These motifs in fact characterize a class of irreducible representations of the Yangian algebra, which span the Fock space of
the HS spin chains.
%In addition, the complete energy spectra of these spin chains along with the %degeneracy factors can be regenerated from the energy functions of some %associated one-dimensional classical vertex models~\cite{BBH10}. 

Interestingly, the Hamiltonian of \su{m|n} HS spin chain may be obtained 
from that of \su{m|n} spin Sutherland model where the particles are equipped with both the dynamical and the spin degrees of freedom~\cite{SS93, Po93}. In the limit of large coupling constant, the spin part of the Hamiltonian of this spin-dynamical model decouples from the dynamical part and reduces to the Hamiltonian of the HS spin chain. This technique of obtaining a spin chain from a spin-dynamical model in the limit of large coupling constant  has a wide range of applicability
for the case of integrable systems. 
Indeed, by using this technique, another spin chain 
with long-range interaction was derived by Polychronakos from the spin Calogero model with confining harmonic potential~\cite{Po93}.
In this case, the lattice sites are inhomogeneously placed on a line and given by the roots of the $N$-th order Hermite Polynomial for $N$ number of sites~\cite{Fr93}. Such a spin chain is known as Polychronakos or Polychronakos-Frahm (PF) spin chain in the literature. The Hamiltonian of the \su{m|n} supersymmetric
version of ferromagnetic PF spin chain with $N$ number of sites is given by~\cite{BUW99} 
\beq
\label{a1}
\mc{H}_N^{(m|n)}=\sum_{1\leq i< j \leq N}
\frac{1-  P_{ij}^{(m|n)}}{(\rho_i-\rho_j)^2}  \, ,
\eeq
where $P_{ij}^{(m|n)}$ is a supersymmetric spin exchange operator which interchanges the spins of the $i$-th and $j$-th lattice sites (along with a phase factor), and $\rho_i$ is the $i$-th root of the $N$-th order Hermite polynomial. It may be noted that, for $n=0$, \eq{a1} simply reduces to the Hamiltonian of the \su{m} ferromagnetic  PF spin chain.  Similar to the case of  \su{m|n} supersymmetric HS spin chain, the PF spin chain \eq{a1} also exhibits $Y(gl_{m|n})$ super Yangian quantum group symmetry for any value of $N$~\cite{HB00}. As a result, 
degenerate multiplets appearing 
in the spectra of these spin chains can again be classified
through the motifs.  
%and reproduce its energy spectrum along with the degeneracy factors from some % one-dimensional classical vertex model~\cite{BBH10}. 

Since the spin and coordinate degrees of freedom of the spin dynamical models 
related to the above mentioned PF and HS spin chains 
are decoupled from each other in the strong coupling limit, the partition functions of these spin chains can be computed by using the so called `freezing trick' \cite{Po94,  BUW99, FG05, BB06}. More precisely, the partition function of such a spin chain 
is obtained 
by taking the ratio of the partition function of the corresponding spin dynamical 
model to that of its spinless version in the strong coupling limit. 
%In the case of \su{m} PF spin chain, Polychronakos computed its
However, it should be noted that, 
the partition functions of the spin chains obtained by using the freezing
trick do not directly lead to the motif representation of the corresponding 
spectra. For this purpose, it is necessary to 
define the so-called generalized partition functions (which reproduce the 
standard partition functions at a certain limit) and to   
apply different techniques for expressing 
those generalized partition functions 
in terms of Schur polynomials associated with the motifs
%or  border strips (which represent the motifs in a graphical way) 
 \cite {KKN97,HB00,BBHS07}. By using 
such expressions of the generalized partition functions,  one can immediately 
identify all degenerate multiplets within the spectra of the corresponding spin chains and  write down the energy eigenvalues related to all motifs. 
Moreover, by applying a rather general framework,
%to such an expression of the partition function, 
it is possible to show that all energy eigenvalues for a  
class of HS like Yangian invariant quantum spin chains can be reproduced from the energy functions of some  
one-dimensional classical vertex models having only local interactions~\cite{BBH10}.
The above mentioned equivalence between the eigenvalues of Yangian invariant
spin chains with long-range interactions and energy functions of  
one-dimensional vertex models with only local interactions
can be extended even in the presence
of chemical potentials~\cite{EFG12,FGLR18}. 
As a result, by using transfer matrices 
associated with those vertex models, one can calculate various thermodynamic quantities  of this type of spin chains even in the presence of chemical potentials  
and investigate the critical properties of those systems~\cite{EFG12,FGLR18,BBCFG19a,BBCFG19b}.
%calculated the canonical partition function of the spin CS model by a particular  
%method which computes the grand canonical partition function at first, expands it 
%in terms of the fugacity parameter and finally extracts the $N$-particle canonical 
%partition function for this spin CS model. 
%can be derived by using the freezing trick \cite{BUW99}.
Thus the expressions of the generalized partition
functions in terms of Schur polynomials lead to a powerful method for 
classifying the degenerate multiplets of the corresponding spectra 
and for studying various thermodynamic properties of the 
related spin chains.   

It may be noted that some homogeneous
multivariate Rogers-Szeg\"o (RS) polynomials, which can be expressed in terms of Schur polynomials, play the role of generalized partition functions for the case of non-supersymmetric PF spin chains 
\cite{Hi95npb, KKN97,Hi95}.  
%re defined as generalized
%partition functions which can be expressed in terms of Schur polynomials. 
Similarly, for the case of \su{m|n} supersymmetric PF spin chain \eq{a1}
with $N$ number of lattice 
sites, one can define a 
multivariate super RS (SRS) polynomial  of the form  \cite{HB00}
\beq
\mbb{H}_{A,N}^{(m|n)}(\mbf{x},\mbf{y};q)= \hskip -.6 cm 
\sum_{\stackrel
{~~~~~a_i ,b_j \, \in \, \mbb{Z}_{\geq 0}}
{~~~~~\sum\limits_{i=1}^{m}a_i+\sum\limits_{j=1}^nb_j=N}}(q)_N\cdot 
q^{\sum\limits_{j=1}^n\frac{b_j(b_j-1)}{2}}
\prod\limits_{i=1}^m
\frac{\mbf{x}_i^{a_i}}{(q)_{a_i}}\prod\limits_{j=1}^n
\frac{\mbf{y}_j^{b_j}}{(q)_{b_j}} \, ,
\label{SRSA}
\eeq
where $\mbb{Z}_{\geq 0}$ represents the set of non-negative integers, 
$\mbf{x}\equiv \mbf{x}_1,\mbf{x}_2,\cdots,\mbf{x}_m$ and
$\mbf{y}\equiv \mbf{y}_1,\mbf{y}_2,\cdots,\mbf{y}_n$ represent
two different sets of variables, $q$ is a free parameter and 
$(q)_n\equiv (1-q)(1-q^2)\cdots (1-q^{n})$.
This SRS polynomial reduces to the partition function of the  supersymmetric PF spin chain  \eq{a1} at temperature $T$ in the  limit $\mbf{x}_1=\mbf{x}_2=\cdots =\mbf{x}_m=1$, 
$\mbf{y}_1=\mbf{y}_2=\cdots=\mbf{y}_n=1$ and for $q= e^{-1/(k_BT)}$.
 Moreover, such SRS polynomials 
corresponding to supersymmetric
PF chains with different numbers of lattice sites ($N$), but 
fixed values of internal degrees of freedom ($m$ and $n$), 
satisfy some recursion relations which lead to the desired expression of   
these polynomials  through super Schur polynomials  \cite{HB00}. 
Hence these SRS polynomials can be treated as generalized partition functions 
for the supersymmetric PF spin chains.

In view of the above discussion, it is natural to ask whether 
multivariate RS or SRS polynomials can be used to analyse the spectra and partition functions of some other PF like spin chains. 
In this context it may be noted that, quantum integrable systems with long-range interactions can be classified according to their connections with  different root systems related to the Lie algebra \cite{OP83,CS02}.  
In particular, the  HS and PF spin chains which have been  
discussed so far are associated with the $A_{N-1}$ type of root system. However, several
exactly solvable variants of the PF spin chain, related to the $BC_N$ and $D_N$ root systems, have also been studied in the 
literature~\cite{YT96,BFGR08,BFGR09,BFG09}. The Hamiltonians of the PF spin chains related to the latter type of root systems contain 
reflection operators like $S_i$ ($i=1,\dots,N$), which satisfy the condition $S_i^2=\one$ and yield a representation of some elements belonging to the $BC_N$ or $D_N$ type of Weyl algebra. As a special case,
$S_i$ can be taken as the spin reversal operator 
$P_i$ which flips the sign of the spin component on the $i$-th lattice site.  
Partition functions of non-supersymmetric $BC_N$ and $D_N$ types of 
PF spin chains, containing such spin reversal operators in their Hamiltonians, 
have been derived by
using the freezing trick \cite{BFGR08,BFG09}. Moreover, by taking 
reflection operators as supersymmetric analogue of  spin reversal operators (SASRO), partition functions of corresponding PF spin chains related to the $BC_N$ root system have also been computed by using the freezing trick \cite{BFGR09}.

However, it is possible to generate wider variants of  $BC_N$ or $D_N$ type of
PF spin chains by 
choosing the reflection operators in more general way than the above mentioned spin reversal operators and their supersymmetric analogues. For example,
in the non-supersymmetric case,
one can choose these reflection operators as arbitrarily polarized spin reversal operators (PSRO) denoted by $P_i^{(m_1,m_2)}$, where $m_1,m_2 \in \mbb{Z}_{\geq 0}$ . This $P_i^{(m_1,m_2)}$ acts as an identity operator
on the first $m_1$ number of elements of the spin basis and as an identity 
operator with a negative sign on the remaining $m_2$ number of elements of the spin basis \cite{BBB14}. It can be shown that, in some particular cases like $m_1=m_2$ or $m_1=m_2\pm1$, $P_i^{(m_1,m_2)}$ reduces to the spin reversal operator $P_i$  through a similarity transformation. Choosing the reflection operators as such PSRO,
new variants of $BC_N$ and $D_N$ type of PF spin chains have been obtained and the partition functions of these spin chains have also been calculated by using the freezing trick~\cite{BBB14,BDFG15}. Finally, by choosing the reflection operators as supersymmetric analogues of PSRO (SAPSRO), an even larger class of $BC_N$  type of PF spin chains have been obtained
~\cite{BBBD16}.  These 
$BC_N$ type of PF spin chains with SAPSRO can generate all of the previously obtained $BC_N$ type of PF spin chains at different limits. The partition 
functions of these $BC_N$  type of PF spin chains with SAPSRO have also been computed by using the freezing trick. Furthermore,  
it has been observed that such partition functions 
can be obtained by taking a certain limit of some 
$BC_N$ type of  multivariate SRS polynomials depending on four different sets of variables \cite{BD17}. 

In spite of the above mentioned works on  $BC_N$  type of PF spin chains, the important problem of classifying the degenerate multiplets of the corresponding spectra through some motif like representations have not been addressed so far.
%Moreover, it is not known whether there exist some one-dimensional vertex models 
%whose energy functions would reproduce the entire spectra of these spin chains. 
The main purpose of the present paper is to solve this problem by using a novel 
expression for  $BC_N$ type of  multivariate SRS polynomials through some 
$q$-deformation of super Schur polynomials.
%express the partition functions of the $BC_N$ type of PF spin chains with SAPSRO
%through some $q$-deformations of Schur polynomials, which would lead to a 
%complete classification of 
%the degenerate multiplets in the corresponding spectra through suitable modifications of the motif    
%representations. 
The arrangement of this paper is as follows. In Section 2 of this paper, we 
define the Hamiltonian of  the $BC_N$ type of ferromagnetic
PF spin chains with SAPSRO and 
briefly summarize some known properties of corresponding multivariate SRS polynomials.  
%By using the Gen function of  these SRS polynomials, 
In Section 3, we show that  these $BC_N$ type of SRS polynomials
satisfy some recursion relations involving a particular type of $q$-deformation  of elementary supersymmetric  polynomials. 
%These recursion relations connect partition functions 
%of $BC_N$ type of PF spin chains with fixed values of internal degrees of freedom  
%and different numbers of lattice sites. 
In Section 4, we use an analogue of  the Jacobi-Trudi formula
to define  a $q$-deformed version of the super Schur polynomials 
in terms of the above mentioned $q$-deformed elementary supersymmetric  polynomials. Subsequently, we expand those $q$-deformed super Schur polynomials
as a power series of the parameter $q$ to obtain the `restricted' super Schur polynomials and also present some combinatorial form of such
restricted super Schur polynomials. In Section 5, we derive  novel  
expressions for the $BC_N$ type of SRS polynomials through  $q$-deformed super Schur polynomials and restricted super Schur polynomials. Such expressions for these  SRS polynomials lead to a  complete classification of 
the degenerate multiplets in the spectra of  $BC_N$ type of ferromagnetic
PF spin chains through the  `branched' motifs. 
For a spin chain with $N$ number of lattice sites, these branched 
motifs may be written as $(\de_1, \de_2,...,\de_{N-1}|l)$, where $\de_i \in \{ 0,1 \}$ and $ l \in \{ 0,1,...,N \}$.  In Section 6, we briefly discuss similar 
classification of 
the degenerate multiplets in the spectra of  $BC_N$ type of anti-ferromagnetic
PF spin chains. In section 7, we use an extended boson-fermion duality 
relation of the restricted super Schur polynomials to show that the partition 
functions of  $BC_N$ type of PF spin chains also exhibit similar
duality relation. Section 8 is the concluding section. 

 %In this paper, we have derived some recursion relations, for the same variants of 
 %RS polynomials, which involve different numbers of lattice sites at fixed values of  %internal degrees of freedom. Using these recursion relations, we have been able 
 %to describe the $BC_N$ type of PF spin chains by `motif' like objects. 

\noi \section{\texorpdfstring{$BC_N $ type of ferromagnetic PF spin chains and SRS polynomials}{BCPF}}
\renewcommand{\theequation}{2.{\arabic{equation}}}
\setcounter{equation}{0}
\medskip

To describe a class of $BC_N$ type of PF spin chains on a superspace, let us consider a set of operators like
$A_{j \alpha}^\dagger$~($B_{j \alpha}$) which creates (annihilates)
a particle of species $\alpha$ on the $j$-th lattice site. Let us assume that these creation (annihilation) operators are bosonic when $\alpha \in [1,2,....,m]$ and 
fermionic when $\alpha \in [m+1,m+2,....,m+n]$.
The parity of these operators are defined as
\bea
 &&\pi(A_{j \alpha})=\pi(A_{j \alpha}^\dagger)=0 ~
\mr{for}~ \alpha \in [1,2,....,m] \, , \nn \\
 &&\pi(A_{j \alpha})=\pi(A_{j \alpha}^\dagger)=1 ~
 \mr{for}~ \alpha \in [m+1,m+2,....,m+n] \, , \nn
\eea
and they satisfy the following commutation (anti-commutation) relations: 
\beq
[A_{j \alpha},A_{k \beta}]_{\pm}=0 \, ,~ 
[A_{j \alpha}^\dagger,A_{k \beta}^\dagger]_{\pm}=0 \, , ~
[A_{j \alpha},A_{k \beta}^\dagger]_{\pm}=\delta_{jk}\delta_{\alpha \beta} \, ,
\label{b2}
\eeq
where $[A,B]_{\pm} \equiv AB- (-1)^{\pi(A)\pi(B)}AB$.
Now, let us consider a finite dimensional
subspace of the corresponding Fock space,  
where each lattice site accommodates only one particle, i.e., 
$\sum_{\alpha=1}^{m+n} A_{j\alpha}^{\dagger} A_{j\alpha}=1$
for all $j\in \{1,2, \cdots, N\}$.  
The supersymmetric spin exchange operator 
$\hat{P}_{ij}^{(m|n)}$ can be defined on this subspace
as \cite{Ha93} 
\beq
\hat{P}_{ij}^{(m|n)} \equiv
\sum_{\alpha,\beta=1}^{m+n} A_{i \alpha}^\dagger
A_{j \beta}^\dagger A_{i \beta}A_{j \alpha} \, . 
\label{b3}
\eeq

The above mentioned supersymmetric spin exchange operator 
can equivalently be expressed as an operator on the total internal space of $N$ number of spins, denoted by $\mbs{\Sigma}^{(m_1,m_2|n_1,n_2)}$, where $m_1,~m_2,~n_1,~n_2 \in \mbb{Z}_{\geq 0}$ 
satisfying the conditions $m_1+m_2=m$ and $n_1+n_2=n$~\cite{Ba99,BBBD16}. 
This $\mbs{\Sigma}^{(m_1,m_2|n_1,n_2)}$ is spanned 
by some orthonormal state vectors of the form  $\ket{s_1,\cdots,
s_i, \cdots, s_N}$, 
where each local spin $s_i \in S\equiv \{1,2,...,m+n\}$ is endowed with two different 
types of parities. For the sake of defining these parities, 
it is convenient to write $S$ as the union of four sets given by 
\beq
\begin{split}
& S_{+,+}^{(m_1)} = \{1,2,...,m_1\}\, , \\
& S_{+,-}^{(m_2)} = \{m_1+1,m_1+2,...,m_1+m_2\}\, , \\
& S_{-,+}^{(n_1)} = \{m_1+m_2+1,m_1+m_2+2,...,m_1+m_2+n_1\}\, , \\
& S_{-,-}^{(n_2)} = \{m_1+m_2+n_1+1,m_1+m_2+n_1+2,...,m_1+m_2+n_1+n_2\}.
\end{split}
\label{sets}
\eeq
 The boson-fermion type parity of the local spin $s_i$ is defined as  
\beq
\begin{split}
 \pi(s_i) & = 0 ~{\rm if}~ s_i \in S_{+,+}^{(m_1)} \cup S_{+,-}^{(m_2)} \, ,\\
& = 1 ~{\rm if}~ s_i \in S_{-,+}^{(n_1)} \cup S_{-,-}^{(n_2)}  \, , 
\end{split}
\eeq
and another parity of  $s_i$, related to the action of SAPSRO, is defined as 
\beq
\begin{split}
 f(s_i) & = 0, ~~{\rm if}~ s_i \in S_{+,+}^{(m_1)} \cup S_{-,+}^{(n_1)} \, ,\\
& = 1, ~~{\rm if}~ s_i \in S_{+,-}^{(m_2)} \cup S_{-,-}^{(n_2)}\, .
\end{split}
\label{SA_parity}
\eeq
Indeed the SAPSRO,   
denoted by $P_i^{(m_1,m_2|n_1,n_2)}$, acts on the basis vectors of the 
space $\mbs{\Sigma}^{(m_1,m_2|n_1,n_2)}$ as~\cite{BBBD16}
\beq
P_i^{(m_1,m_2|n_1,n_2)}\ket{s_1,\cdots,s_i,\cdots, s_N}
=(-1)^{f(s_i)}\ket{s_1,\cdots,s_i,\cdots, s_N}.
\label{b7}
\eeq
Since each local spin vector $s_i$ may be chosen in $(m+n)$ number of 
different ways, 
 $\mbs{\Sigma}^{(m_1,m_2|n_1,n_2)}$ can be expressed as a direct product of the form
\beq
\mbs{\Sigma}^{(m_1,m_2|n_1,n_2)} 
\equiv \underbrace{\mc{C}_{m+n} 
\otimes \mc{C}_{m+n} 
\otimes \cdots \otimes \mc{C}_{m+n}}_{N} \, ,  
\label{b5}
\eeq
where $\mc{C}_{m+n}$ is an $(m+n)$-dimensional complex vector space.
Hence 
%it is clear that 
$\mbs{\Sigma}^{(m_1,m_2|n_1,n_2)} $
is isomorphic to the subspace of the Fock space,
on which $\hat{P}_{ij}^{(m|n)}$ in \eq{b3} is defined.
A supersymmetric spin exchange operator $P_{ij}^{(m|n)}$ can be defined on the space $\mbs{\Sigma}^{(m_1,m_2|n_1,n_2)}$ as~\cite{BB06,Ba99} 
\beq
P_{ij}^{(m|n)}\ket{s_1,\cdots,s_i,\cdots,s_j,\cdots,s_N}
=(-1)^{\alpha_{ij}(\mbf{s})}
\ket{s_1,\cdots,s_j,\cdots,s_i,\cdots,s_N},
\label{b6}
\eeq
where
$\alpha_{ij}(\mbf{s})
=\pi(s_i)\pi(s_j)+\left(\pi(s_i)+\pi(s_j)\right)\, 
\rho_{ij}(\mbf{s})$
and
$\rho_{ij}(\mbf{s})=\sum_{k=i+1}^{j-1}\pi(s_k)$. 
The above equation implies that if two spins $s_i$ and $s_j$ with $\pi(s_i)=\pi(s_j)=0$ or $\pi(s_i)=\pi(s_j)=1$ are exchanged, then one gets a phase factor of $1$ or $-1$ respectively.
Hence $s_i$ can be considered as a `bosonic' spin  if 
$\pi(s_i)=0$ and a `fermionic' spin if
$\pi(s_i)=1$. However, it must be noted that the exchange of a bosonic spin with a fermionic spin (or, vice versa) produces a nontrivial phase factor of $(-1)^{\rho_{ij}(\mbf{s})}$, where $\rho_{ij}(\mbf{s})$ represents the number of fermionic spins between the $i$-th and $j$-th lattice sites. 
Applying the commutation (anti-commutation) relations given in \eq{b2}, 
one can easily  show that $\hat{P}_{ij}^{(m|n)}$
in \eq{b3} is completely equivalent to  
$P_{ij}^{(m|n)}$ in \eq{b6}. 

It may be noted  that,  $P_{ij}^{(m|n)}$ in \eq{b6}  
and $P_i^{(m_1,m_2|n_1,n_2)}$ in \eq{b7}  
yield a representation 
of the $BC_N$ type of Weyl algebra~\cite{BBBD16}.
Using this representation of Weyl algebra, 
the Hamiltonian of a class of exactly solvable $BC_N$
type of ferromagnetic PF spin chains has been defined in the latter reference 
as
\beq
\mc{H}^{(m_1,m_2|n_1,n_2)}_N
=\sum_{\underset{i \neq j}{i,j=1}}^N \left[\frac{1- P_{ij}^{(m|n)}}{(\xi_i-\xi_j)^2} +
\frac{1- \widetilde{{P}}_{ij}^{(m_1,m_2|n_1,n_2)}}{(\xi_i+\xi_j)^2}\right]
+\beta\sum_{i=1}^{N}\frac{1- P_i^{(m_1,m_2|n_1,n_2)} }{\xi_i^2} \, ,  
\label{b8}
\eeq
where 
 $\beta>0$ is a real parameter,
 $\xi_i=\sqrt{2y_i}$ with $y_i$ 
being the $i$-th root  
of the generalized Laguerre polynomial $L_N^{\beta -1}$ and 
$\widetilde{P}_{ij}^{(m_1,m_2|n_1,n_2)} \equiv 
 P_i^{(m_1,m_2|n_1,n_2)}P_j^{(m_1,m_2|n_1,n_2)} P_{ij}^{(m|n)}$.
The above mentioned Hamiltonian 
is able to reproduce all of the previously studied $BC_N$ type of
PF spin chains 
for different values of the discrete parameters 
$m_1,~m_2,~n_1$ and $n_2$. For example, in the case when all the spins are either bosonic or fermionic, i.e., either $n_1=n_2=0$ or $m_1=m_2=0$, 
$\mc{H}^{(m_1,m_2|n_1,n_2)}_N$ given by \eq{b8}
reduces to the non-supersymmetric PF spin chain 
containing PSRO~\cite{BBB14}. 
Moreover, in another case, where  
the discrete parameters in \eq{b8}
%of $\mc{H}^{(m_1,m_2|n_1,n_2)}_N$ 
satisfy the following relations:
\beq
m_1= \frac{1}{2}\left( m+ \ep \, \tilde{m} \right),~ 
m_2= \frac{1}{2}\left( m - \ep \, \tilde{m} \right),~
n_1= \frac{1}{2}\left( n+ \ep' \, \tilde{n} \right), ~
n_2= \frac{1}{2}\left( n - \ep' \, \tilde{n} \right),
\label{con}
\eeq
with $\ep,\ep^{\prime}=\pm 1$, $\tilde{m}
\equiv m~ \mr{mod}~ 2$ and
$\tilde{n} \equiv n~ \mr{mod}~ 2$,
%it is possible to give a unitary transformation to 
one can obtain
the Hamiltonian (depending 
on the parameters $m,n,\ep,\ep'$)
%$\mc{H}_N^{(m|n)}_{\ep \ep'}$ 
of the $BC_N$ type of 
PF spin chains with SASRO ~\cite{BFGR09}
by using a unitary transformation~\cite{BBBD16}.

 Expanding the grand canonical partition function of the corresponding spin Calogero model as a power series of the fugacity parameter and 
 applying the freezing trick, the canonical partition function of the $BC_N$
 type of ferromagnetic PF spin chain \eq{b8} has been derived  
 in the form~\cite{BD17}
 \beq
\mc{Z}^{(m_1,m_2|n_1,n_2)}_{B,N}(q) 
= %\hskip -.2 cm 
\sum_{\stackrel
{a_i, \, b_j, \, c_k, \,  d_l \, \in \, \Z}
{\sum\limits_{i=1}^{m_1} a_i+\sum\limits_{j=1}^{m_2} b_j
+\sum\limits_{k=1}^{n_1} c_k+\sum\limits_{l=1}^{n_2} d_l=N}
}~
\frac{\left(q^2\right)_N \cdot q^{\sum\limits_{j=1}^{m_2} b_j+\sum\limits_{k=1}^{n_1} c_k(c_k-1)+\sum\limits_{l=1}^{n_2} d_l^2}}
{\prod\limits_{i=1}^{m_1}(q^2)_{a_i}\prod\limits_{j=1}^{m_2}(q^2)_{b_j}\prod\limits_{k=1}^{n_1}(q^2)_{c_k}
\prod\limits_{l=1}^{n_2}(q^2)_{d_l}} \, ,
\label{b9}
\eeq 
where $q\equiv e^{-1/(k_BT)}$ and, for the sake of convenience,
the above partition function is defined as a function of $q$ instead of $T$. 
It may be noted that this partition function does not depend
on the 
parameter $\beta$ which is present in the Hamiltonian \eq{b8}.
Motivated by the form of this partition function, a class of 
$BC_N$ type of SRS polynomials has  also been introduced  
in the later reference as 
\bea
&& \mbb{H}_{B,N}^{(m_1,m_2|n_1,n_2)}(x,\bar x;y,\bar y;q)\nn \\
&&= \hskip -.2 cm \sum_{\stackrel
{a_i, \, b_j, \, c_k, \,  d_l \, \in \, \Z}
{\sum\limits_{i=1}^{m_1} a_i+\sum\limits_{j=1}^{m_2} b_j
+\sum\limits_{k=1}^{n_1} c_k+\sum\limits_{l=1}^{n_2} d_l=N}
} 
\hskip -1.64 cm (q^2)_N \cdot 
q^{\sum\limits_{j=1}^{m_2} b_j+\sum\limits_{k=1}^{n_1} c_k(c_k-1)+\sum\limits_{l=1}^{n_2} d_l^2}  \cdot 
\prod\limits_{i=1}^{m_1}\frac{x_i^{a_i}}{(q^2)_{a_i}}
\prod\limits_{j=1}^{m_2}\frac{(\bar{x}_j)^{b_j}}{(q^2)_{b_j}}
\prod\limits_{k=1}^{n_1}\frac{y_k^{c_k}}{(q^2)_{c_k}}
\prod\limits_{l=1}^{n_2}\frac{(\bar{y}_l)^{d_l}}{(q^2)_{d_l}}
\, , \nn \\
&&~~~~
\label{SRS0}
\eea
(with $\mbb{H}_{B,0}^{(m_1,m_2|n_1,n_2)}(x,\bar x,y,\bar y;q)=1$),
where $x\equiv x_1,x_2,\cdots,x_{m_1}$, 
$\bar x\equiv \bar x_1, \bar x_2,\cdots, \bar x_{m_2}$, 
$y\equiv y_1,y_2,\cdots,y_{n_1}$ and 
$\bar y\equiv \bar y_1, \bar y_2,\cdots, \bar y_{n_2}$
denote four different sets of variables. It is evident that the partition function \eq{b9} can be obtained by taking a limit of the SRS polynomial \eq{SRS0}
 as
\beq
\mc{Z}_{B,N}^{(m_1,m_2|n_1,n_2)}(q)
=\mbb{H}_{B,N}^{(m_1,m_2|n_1,n_2)}(x=1,\bar{x}=1,y=1,\bar{y}=1;q).
\label{SRSpa}
\eeq
The generating function corresponding to the $BC_N$ type of SRS 
polynomials \eq{SRS0} has been defined as \cite{BD17}
\beq
\mc{G}_B^{(m_1,m_2|n_1,n_2)}(x,\bar{x};y,\bar{y};q,t)=\mc{G}_1^{(m_1)}(x;q,t) \cdot \mc{G}_2^{(m_2)}(\bar{x};q,t) \cdot 
\mc{G}_3^{(n_1)}(y;q,t) \cdot  
\mc{G}_4^{(n_2)}(\bar{y};q,t) \, ,
\label{GF}
\eeq
where 
\begin{subequations}
\bea
&&\mc{G}_1^{(m_1)}(x;q,t)=\frac{1}{\prod\limits_{i=1}^{m_1}(tx_i;q^2)_{\infty}} \, , 
\label{g1} \\
&&\mc{G}_2^{(m_2)}(\bar{x};q,t)=\frac{1}{\prod\limits_{j=1}^{m_2}(tq\bar{x}_j;q^2)_{\infty}} \, , 
\label{g2} \\
&&\mc{G}_3^{(n_1)}(y;q,t)=\frac{1}{\prod\limits_{k=1}^{n_1}(-tq^{-2}y_k;q^{-2})_{\infty}} \, , \label{g3} \\
&&\mc{G}_4^{(n_2)}(\bar{y};q,t)=\frac{1}{\prod\limits_{l=1}^{n_2}(-tq^{-1}\bar{y}_l;q^{-2})_{\infty}} \, ,  \label{g4}
\eea
\label{GF2}
\end{subequations}
and the notation
$ (t;q)_\infty \equiv \prod\limits_{i=1}^{\infty} (1-tq^{i-1}) $  
%(and $(t;q)_0=1$) 
has been used. Expanding the generating function \eq{GF}  as a power
series of the parameter $t$
by using the identity  \cite{An76} 
\bea
\frac{1}{(t;q)_{\infty}}=\sum\limits_{N=0}^{\infty}
\, \frac{t^N}{(q)_N} \, , 
\label{sumid}
\eea
one can show that  
\beq
\mc{G}_B^{(m_1,m_2|n_1,n_2)}(x,\bar{x};y,\bar{y};q,t)=\sum\limits_{N=0}^{\infty}\frac{
\mbb{H}_{B,N}^{(m_1,m_2|n_1,n_2)}(x,\bar x;y,\bar y;q)}
{(q^2)_N} ~ t^N  .
\label{Gen}
\eeq
Using this generating function, some recursion relations for 
the $BC_N$ type of SRS 
polynomials associated with different number of lattice sites and internal degrees of freedom have  been found in Ref.~\cite{BD17}. However, one can not use
%it is easy to check that
such recursion relations for expressing 
the $BC_N$ type of SRS polynomials through some  super Schur like polynomials. So, in the next section, we shall derive a different type of recursion relations for the $BC_N$ type of SRS polynomials involving different number of lattice sites ($N$), but with 
fixed values of the internal degrees of freedom $(m_1,m_2,n_1,n_2)$. This later type of recursion relation will enable us to express the $BC_N$ type of SRS polynomials through some $q$-deformation of  super Schur polynomials.

%play a crucial role in finding the `motif' like representation of the $BC_N$ type of 
%PF spin chain. 

\noi \section{\texorpdfstring{Novel recursion relations for $BC_N $ type of SRS polynomials}{BCPFrecursion}}
\renewcommand{\theequation}{3.{\arabic{equation}}}
\setcounter{equation}{0}
\medskip
For the purpose of deriving recursion relations for the $BC_N$ type of SRS polynomials, involving different number of lattice sites at fixed values of internal degrees of freedom, we may proceed in the following way. By using \eq{g1},  we 
obtain
\begin{equation*}
\begin{split}
\mc{G}_1^{(m_1)}(x;q,t)
    &= \frac{1}{\prod\limits_{i=1}^{m_1} \{ (1-tx_i)(1-q^2tx_i)(1-q^4tx_i)\cdots \}}\\
    &= \frac{1}{\prod\limits_{i=1}^{m_1}  (1-tx_i) \prod\limits_{i=1}^{m_1} 
    (q^2tx_i;q^2)_{\infty} \,  }
    \, ,
\end{split} 
\end{equation*}
which implies that
\beq
\label{g1q2}
 \mc{G}_1^{(m_1)}(x;q,q^2t)= 
 {\prod\limits_{i=1}^{m_1} (1-tx_i)} \cdot \mc{G}_1^{(m_1)}(x;q,t).
\eeq
Similarly, by using \eq{g2}, one can show that  
\bea
\label{g2q2}
 \mc{G}_2^{(m_2)}(\bar{x};q,q^2 t)= {\prod\limits_{j=1}^{m_2} (1-tq\bar{x_j})} \cdot  \mc{G}_2^{(m_2)}(\bar{x};q, t)
\eea
Next, by using \eq{g3},  we obtain 
\[
\begin{split}
    \mc{G}_3^{(n_1)}(y;q,t) 
    &= \frac{1}{\prod\limits_{k=1}^{n_1} \{ (1+ty_kq^{-2})(1+ty_kq^{-4})
    \cdots \}}\\
    &= \frac{\prod\limits_{k=1}^{n_1} (1+ty_k)}{\prod\limits_{k=1}^{n_1} \{ 
    (1+ty_k)(1+ty_kq^{-2})(1+ty_kq^{-4})\cdots \}} \, ,
   % = \frac{\prod\limits_{k=1}^{n_1} (1+ty_k)}{\prod\limits_{k=1}^{n_1} 
   %(-ty_k;q^{-2})_{\infty}} \, ,
\end{split}
\]
which implies that 
\beq
\label{g3q2}
\mc{G}_3^{(n_1)}(y;q,q^2t) = \prod\limits_{k=1}^{n_1} \frac{1}{(1+ty_k)}~ \mc{G}_3^{(n_1)}(y;q,t).
\eeq
Similarly, by using \eq{g4}, we find that 
\beq
\label{g4q2}
\mc{G}_4^{(n_2)}(\bar{y};q,q^2 t) = \frac{1}{\prod\limits_{l=1}^{n_2} (1+tq\bar{y_l})}~ \mc{G}_4^{(n_2)}(\bar{y};q,t).
\eeq
Combining Eqs. \eq{g1q2}, \eq{g2q2}, \eq{g3q2} and \eq{g4q2}, and using the definition of the generating function \eq{GF}, we obtain the following $q^2$-difference relation 
\beq
\label{GFq2}
\mc{G}_B^{(m_1,m_2|n_1,n_2)}(x,\bar x;y,\bar y;q,q^2t)=\frac{\prod\limits_{i=1}^{m_1}(1-tx_i) \prod\limits_{j=1}^{m_2}(1-tq\bar{x}_j)}{\prod\limits_{k=1}^{n_1}(1+ty_k) \prod\limits_{j=1}^{n_2}(1+tq\bar{y}_l)}~\mc{G}_B^{(m_1,m_2|n_1,n_2)}(x,\bar x;y,\bar y;q,t).
\eeq

In this context it may be noted that, 
%for the case of \su{m|n} PF spin chains \eq{a1}, 
the generating function $\mc{G}_A^{(m|n)}(\mbf{x},\mbf{y};q,t)$ for the $A_{N-1}$ type of SRS polynomials satisfies a similar $q$-difference relation given by \cite{HB00}
\beq
\label{GFq0}
\mc{G}_A^{(m|n)}(\mbf{x},\mbf{y};q,qt)=\frac{\prod\limits_{i=1}^{m}(1-t\mbf{x}_i)}{\prod\limits_{j=1}^{n}(1+t\mbf{y}_j) }~\mc{G}_A^{(m|n)}(\mbf{x},\mbf{y};q,t).
\eeq
The product $\prod\limits_{i=1}^{m}(1-t\mbf{x}_i)$ appearing in the above equation
can be expanded as a polynomial in $t$ as 
%\beq
\begin{align}
   \prod\limits_{i=1}^{m}(1-t\mbf{x}_i) 
   %&= (1+tx_1)(1+tx_2) \cdots (1+tx_m) \nn \\
  %&= 1 + t\sum\limits_{i=1}^m x_i + t^2\sum\limits_{i<j}^m x_ix_j + \cdots 
  %+ t^r\sum\limits_{\al_1 < \al_2 < \cdots < \al_r} x_{\al_1} x_{\al_2} \cdots 
  %x_{\al_r} \nn \\
  %& ~~~~~~~~~~~~~~~~~ + \cdots + t^m\sum\limits_{\al_1 < \al_2 <  
  %\cdots < \al_m} x_{\al_1} x_{\al_2} \cdots x_{\al_m} \nn \\
  &= \sum_{r=0}^{m} \, (-1)^r  \, t^r \, e_r^{(m)}(\mbf{x}) \, ,
  \label{elgen}
\end{align}
  where
\[
%\label{ele_pol}
e_r^{(m)}(\mbf{x}) \equiv 
% e_r^{(m)}(x_1,x_2,...,x_m) 
\sum_{1 \leq \al_1 < \al_2 < \cdots < \al_r \leq m} \mbf{x}_{\al_1} \mbf{x}_{\al_2}
\cdots \mbf{x}_{\al_r} \, ,
\]
and  $e_0^{(m)}(\mbf{x})=1$. This $e_r^{(m)}(\mbf{x})$
is known as the elementary symmetric polynomial of $m$-variables 
($\mbf{x}_1,...,\mbf{x}_m$) with degree $r$,
which vanishes for $r>m$. For the special case $\mbf{x}=1$, this polynomial reduces to 
\beq
\label{elval}
e_r^{(m)}(\mbf{x})|_{\mbf{x}=1} = C^m_r \, ,
\eeq
with $C^m_r$ being the binomial coefficient. 
Similarly, the product $\prod\limits_{j=1}^{n} \frac{1}{(1+t\mbf{y}_j)}$ 
can be expanded as an infinite power series of  $t$ as 
\begin{align}
 \prod\limits_{j=1}^{n} \frac{1}{(1+t\mbf{y}_j)} 
 %&= \prod\limits_{i=1}^{n} \sum\limits_{l_i=0}^{\infty} (ty_i)^{l_i} 
   % &= \sum\limits_{K_1,...,K_n=0}^{\infty} t^{\sum\limits_{i=1}^{n}} \prod
   %\limits_{i=1}^{n} y_i^{K_i} \nn \\
    %&= \sum\limits_{r=0}^{\infty} t^r \sum\limits_{K_1+...+K_n=r} \prod
    %\limits_{i=1}^{n} y_i^{K_i} \nn \\
    = \sum_{r=0}^{\infty} (-1)^r \, t^r \, h_r^{(n)}(\mbf{y}) \, ,
\label{comgen}   
\end{align}
where 
\[
h_r^{(n)}(\mbf{y}) \equiv \sum\limits_{\stackrel{l_1+...+l_n=r}{l_1,\cdots , l_n
\in \Z}} \prod\limits_{i=1}^{n} \mbf{y}_i^{l_i}
\]
(and $h_0^{(n)}(\mbf{y})=1$)
is known as the completely symmetric polynomial of $n$-variables with degree $r$. 
It may be noted that 
\bea
\label{comval}
h_r^{(n)}(\mbf{y})|_{y=1} = C^{n+r-1}_r \, ,
\eea
which is non-zero for any non-negative value of $r$.
Combining Eqs. \eq{elgen} and \eq{comgen}, one obtains  
\beq
\frac{\prod\limits_{i=1}^m (1-t\mbf{x}_i)}{\prod\limits_{j=1}^n (1+t\mbf{y}_j)}
= \sum\limits_{k=0}^{\infty} (-1)^k \, t^k \, E_k^{(m|n)}(\mbf{x};\mbf{y})\, ,
\label{prodex}
\eeq
where the elementary supersymmetric  polynomial
$E_k^{(m|n)}(\mbf{x};\mbf{y})$ is 
defined as  
\beq
E_k^{(m|n)}(\mbf{x};\mbf{y}) \equiv \sum\limits_{l=0}^k e_l^{(m)}(\mbf{x})~ h_{k-l}^{(n)}(\mbf{y}).
\label{elsuA}
\eeq
Substituting \eq{prodex}  into \Eq{GFq0}, 
and also using the expansion of 
$\mc{G}_A^{(m|n)}(\mbf{x},\mbf{y};q,t)$ in terms 
of the corresponding SRS polynomials \eq{SRSA}, it has been found that these 
polynomials satisfy a recursion relation of the form \cite{HB00}
\beq
\mbb{H}_{A,N}^{(m|n)}(\mbf{x},\mbf{y};q)=\sum\limits_{k=1}^N (-1)^{k+1} \frac{(q)_{N-1}}{(q)_{N-k}} \cdot E_k^{(m|n)}(\mbf{x};\mbf{y}) \cdot \mbb{H}_{A,N-k}^{(m|n)}(\mbf{x},\mbf{y};q) \, .
\label{recur}
\eeq

At present, our goal is to find out an analogue of the recursion relation \eq{recur}
for the case of  $BC_N$ type of SRS polynomials. To this end, 
we use the expansion \eq{elgen} to obtain 
%\begin{equation*}
  %  \begin{split}
   % \prod\limits_{i=1}^{m_1} (1-tx_i) \prod\limits_{j=1}^{m_2} (1-tq
   %\bar{x_j}) &= \sum\limits_{r=0}^{m_1} \sum\limits_{s=0}^{m_2} (-t)^{r
   %+s}~ q^s~ e_r^{(m_1)}(x)~ e_s^{(m_2)}(\bar{x})\\
    %&= \sum\limits_{k=0}^{m_1+m_2} (-1)^k~ t^k ~\sum\limits_{r=0}^k 
    %q^{k-r}~ e_r^{(m_1)}(x)~ e_{k-r}^{(m_2)}(\bar{x}).
%\end{split}
%\end{equation*}
%Hence we can write 
\beq
\label{qdefel}
\prod\limits_{i=1}^{m_1} (1-tx_i) \prod\limits_{j=1}^{m_2} (1-tq\bar{x}_j) = \sum\limits_{k=0}^{m_1+m_2} (-1)^k~ t^k~ e_k^{(m_1,m_2)}(x,\bar{x};q) \, ,
\eeq
where $e_k^{(m_1,m_2)}(x,\bar{x};q)$ is 
defined as 
\beq
\label{elpolq}
e_k^{(m_1,m_2)} (x,\bar{x};q) \equiv \sum\limits_{r=0}^k q^{k-r}~ e_r^{(m_1)}(x)~ e_{k-r}^{(m_2)}(\bar{x}).
\eeq
Hence, $e_k^{(m_1,m_2)}(x,\bar{x};q)$ is a homogeneous polynomial of degree $k$ of the variables $x$ and $\bar{x}$, and this polynomial 
vanishes for $k>m_1+m_2$. It may be noted that,
for the particular case $q=1$,  \eq{qdefel} reduces to 
\beq
\prod\limits_{i=1}^{m_1} (1-tx_i) \prod\limits_{j=1}^{m_2} (1-t\bar{x}_j) = \sum\limits_{k=0}^{m_1+m_2} (-1)^k ~t^k~ e_k^{(m_1,m_2)}(x,\bar{x};q=1).
\eeq
Comparing this equation with the direct expansion of $\prod\limits_{i=1}^{m_1} (1-tx_i) \prod\limits_{j=1}^{m_2} (1-t\bar{x}_j)$ by using 
\eq{elgen}, we find that 
\beq
\label{eq1}
e_k^{(m_1,m_2)}(x,\bar{x};q=1) = e_k^{(m_1+m_2)}(x,\bar{x}) \, ,
\eeq
where 
$e_k^{(m_1+m_2)}(x,\bar{x}) \equiv e_k^{(m)}(\mbf{x}) $, with $\mbf{x} \equiv x, \bar{x}$.
Hence $e_k^{(m_1,m_2)}(x,\bar{x};q)$  in \eq{elpolq} may be considered as a particular type of  
$q$-deformation of the elementary symmetric polynomial $e_k^{(m_1+m_2)}(x,\bar{x})$. 
Moreover, since $e_{k-r}^{(m_2)}(\bar{x})$ is a homogeneous polynomial of degree $k-r$, \Eq{elpolq} can also be expressed as
\beq
\label{x_qx}
e_k^{(m_1,m_2)}(x,\bar{x};q)= \sum\limits_{r=0}^k e_r^{(m_1)}(x) e_{k-r}^{(m_2)}(q\bar{x}) = e_k^{(m_1+m_2)}(x,\bar{x})|_{\bar{x} \rightarrow q\bar{x}}\, .
\eeq

Next, by using the expansion \eq{comgen}, we similarly find that 
\beq
\label{q_def_com}
\frac{1}{\prod\limits_{k=1}^{n_1}(1+ty_k) \prod\limits_{j=1}^{n_2}(1+tq\bar{y}_l)} = \sum\limits_{k=0}^{\infty} (-1)^k \, t^k \,
h_k^{(n_1,n_2)} (y,\bar{y};q) \, ,
\eeq
where $h_k^{(n_1,n_2)} (y,\bar{y};q) $ is defined as 
\beq
\label{com_pol_q}
h_k^{(n_1,n_2)} (y,\bar{y};q) \equiv \sum\limits_{r=0}^k q^{k-r} \,  h_r^{(n_1)}(y) \,  h_{k-r}^{(n_2)}(\bar{y})\, .
\eeq
Moreover, for the case $q=1$, the above equation yields 
\beq
\label{hq1}
h_k^{(n_1,n_2)}(y,\bar{y};q=1) = h_k^{(n_1+n_2)}(y,\bar{y}) \, ,
\eeq
where $h_k^{(n_1+n_2)}(y,\bar{y}) \equiv h_k^{(n)}(\mbf{y})$, with $\mbf{y} \equiv y,\bar{y}$.
Hence $h_k^{(n_1,n_2)} (y,\bar{y};q)$ in \eq{com_pol_q} is a homogeneous polynomial of degree k of the variables $y$ and $\bar{y}$, which may be considered as a   
$q$-deformation of the completely symmetric polynomial
$h_k^{(n_1+n_2)}(y,\bar{y})$.
Also, in analogy with the case of $q$-deformed elementary symmetric
polynomials, we can rewrite this $h_k^{(n_1,n_2)}(y,\bar{y};q)$ as 
\bea
\label{y_qy}
h_k^{(n_1,n_2)}(y,\bar{y};q)= \sum\limits_{r=0}^k h_r^{(n_1)}(y)~ h_{k-r}^{(n_2)}(q\bar{y}) = h_k^{(n_1+n_2)}(y,\bar{y})|_{\bar{y} \rightarrow q\bar{y}} \, .
\eea
%It is evident from \eq{eq1} and \eq{hq1} that $e_k^{(m_1,m_2)} (x,
%\bar{x};q)$ and $h_k^{(n_1,n_2)} (y,\bar{y};q)$ can be regarded as $q$-%deformations of $e_k^{(m_1+m_2)}(x,\bar{x})$ and $h_k^{(n_1+n_2)}(y,%
%\bar{y})$ respectively. 

Combining Eqs. \eq{qdefel} and \eq{q_def_com}, we obtain  
\beq
 \label{factor}
 \frac{\prod\limits_{i=1}^{m_1}(1-tx_i) \prod\limits_{j=1}^{m_2}(1-tq\bar{x}_j)}{\prod\limits_{k=1}^{n_1}(1+ty_k) \prod\limits_{l=1}^{n_2}(1+tq\bar{y}_l)} = \sum\limits_{k=0}^{\infty} (-1)^k~ t^k~ E_k^{(m_1,m_2|n_1,n_2)}(x,\bar{x};y,\bar{y};q) \, ,
\eeq
where  $E_k^{(m_1,m_2|n_1,n_2)}(x,\bar{x};y,\bar{y};q) $ is defined as 
 \beq
\label{super_el_q}
E_k^{(m_1,m_2|n_1,n_2)}(x,\bar{x};y,\bar{y};q) \equiv \sum\limits_{k_1=0}^{k} e_{k_1}^{(m_1,m_2)}(x,\bar{x};q) \, h_{k-k_1}^{(n_1,n_2)}(y,\bar{y};q).
\eeq
Putting $q=1$ in the above equation and also using 
Eqs.\eq{eq1}, \eq{hq1} and \eq{elsuA} consecutively, we find that
\bea
E_k^{(m_1,m_2|n_1,n_2)}(x,\bar{x};y,\bar{y};q=1) &=& \sum\limits_{k_1=0}^{k} e_{k_1}^{(m_1+m_2)}(x,\bar{x})\,  h_{k-k_1}^{(n_1+n_2)}(y,\bar{y}) \nn \\
&=& E_k^{(m_1+m_2|n_1+n_2)}(x,\bar{x};y,\bar{y}) \, ,
\label{E_q_1}
\eea
where $E_k^{(m_1+m_2|n_1+n_2)}(x,\bar{x};y,\bar{y}) \equiv E_k^{(m|n)}(\mbf{x};\mbf{y})$, 
with $\mbf{x}\equiv x, \bar{x}$ and $\mbf{y}\equiv y, \bar{y}$. 
Thus $E_k^{(m_1,m_2|n_1,n_2)}(x,\bar{x};y,\bar{y};q)$
may be seen as a $q$-deformation of the elementary supersymmetric  polynomial
$E_k^{(m_1+m_2|n_1+n_2)}(x,\bar{x};y,\bar{y})$.
Moreover, by using Eqs.~\eq{super_el_q}, \eq{x_qx}, 
\eq{y_qy} and \eq{elsuA} consecutively, we find that 
$E_k^{(m_1,m_2|n_1,n_2)}(x,\bar{x};y,\bar{y};q)$ can be obtained from 
$E_k^{(m_1+m_2|n_1+n_2)}(x,\bar{x};y,\bar{y})$
by scaling some of its variables as 
\beq
\label{x_qx_y_qy}
E_k^{(m_1,m_2|n_1,n_2)}(x,\bar{x};y,\bar{y};q) = E_k^{(m_1+m_2|n_1+n_2)}(x,\bar{x};y,\bar{y})|_{\bar{x} \rightarrow q\bar{x}, \bar{y} \rightarrow q\bar{y}}.
\eeq

Subsequently, expanding both sides of \Eq{GFq2} in powers of $t$ by using  Eqs. \eq{Gen} and \eq{factor}, we get
\begin{equation*}
\begin{split}
& \sum\limits_{N=0}^{\infty} (q^2t)^N  ~\frac{
\mbb{H}_{B,N}^{(m_1,m_2|n_1,n_2)}(x,\bar x;y,\bar y;q)}
{(q^2)_N}\\
&= \sum\limits_{k=0}^{\infty} \sum\limits_{s=0}^{\infty} (-1)^k \, t^{k+s} \,  \frac{1}{(q^2)_s} \, E_k^{(m_1,m_2|n_1,n_2)}(x,\bar{x};y,\bar{y};q) \, 
\mbb{H}_{B,s}^{(m_1,m_2|n_1,n_2)}(x,\bar x,y,\bar y;q).
\end{split}
\end{equation*}
Redefining the variable $k$ as $k=N-s$, one can rewrite the above equation as 
\beq
\begin{split}
& \sum\limits_{N=0}^{\infty} (q^2t)^N~ \frac{
\mbb{H}_{B,N}^{(m_1,m_2|n_1,n_2)}(x,\bar x;y,\bar y;q)}
{(q^2)_N}\\
&= \sum\limits_{N=0}^{\infty} t^N \sum\limits_{k=0}^{N} (-1)^k ~ \frac{1}{(q^2)_{N-k}}~ E_k^{(m_1,m_2|n_1,n_2)}(x,\bar{x};y,\bar{y};q)~ \mbb{H}_{B,N-k}^{(m_1,m_2|n_1,n_2)}(x,\bar x;y,\bar y;q).
\end{split}
 \eeq
 Comparing the powers of $t^N$ from both sides of the above equation, we finally obtain the following recursion relation  
\bea
\label{RS_recursion}
&& \mbb{H}_{B,N}^{(m_1,m_2|n_1,n_2)}(x,\bar x;y,\bar y;q)\nn \\
&&= \sum\limits_{k=1}^N (-1)^{k-1}~ \frac{(q^2)_{N-1}}{(q^2)_{N-k}}~ E_k^{(m_1,m_2|n_1,n_2)}(x,\bar{x};y,\bar{y};q)~ \mbb{H}_{B,N-k}^{(m_1,m_2|n_1,n_2)}(x,\bar x;y,\bar y;q) \, ,
~~~~~~
\eea
which involves $BC_N$ type of SRS polynomials associated with  
different number of lattice sites and 
fixed values of the internal degrees of freedom. % $(m_1,m_2,n_1,n_2)$. 
It may be noted that, 
the above  relation can generate any $\mbb{H}_{B,N}^{(m_1,m_2|n_1,n_2)}(x,\bar x;y,\bar y;q)$ in a recursive way from the given initial condition $\mbb{H}_{B,0}^{(m_1,m_2|n_1,n_2)}(x,\bar x;y,\bar y;q)=1$. The first few SRS polynomials of $BC_N$ type are found to be
\begin{align}
   \label{RS1}
    & \mbb{H}_{B,1}(q) = E_1(q)\, , \\
    \label{RS2}
    & \mbb{H}_{B,2}(q) = \left(E_1(q)\right)^2-(1-q^2)E_2(q)\, , \\
    \label{RS3}
    & \mbb{H}_{B,3}(q) = \left(E_1(q)\right)^3+(q^2-1)(q^2+2)E_1(q)E_2(q)+(1-q^2)(1-q^4)E_3(q) \, , 
\end{align}
where we have used the following shorthand   notation:
\beq
%\label{RS_notation}
    ~~~~~~~~\mbb{H}_{B,N}(q) \equiv \mbb{H}_{B,N}^{(m_1,m_2|n_1,n_2)}(x,\bar x;y,\bar y;q) \, ,~~E_k(q) \equiv E_k^{(m_1,m_2|n_1,n_2)}(x,\bar{x};y,\bar{y};q) \, .
    \nn ~~~~(3.32a,b)
\eeq
\addtocounter{equation}{1}
By putting $x=1,\bar{x}=1,y=1,\bar{y}=1$ in the recursion relation \eq{RS_recursion} and then by using \eq{SRSpa}, we can also write a similar recursion relation in terms of the partition function of the $BC_N$ type of PF spin chain as 
\beq
\begin{split}
& \mc{Z}_{B,N}^{(m_1,m_2|n_1,n_2)}(q)\\
& = \sum\limits_{k=1}^N (-1)^{k-1}~ \frac{(q^2)_{N-1}}{(q^2)_{N-k}}~ E_k^{(m_1,m_2|n_1,n_2)}(x=1,\bar{x}=1,y=1,\bar{y}=1;q)~ \mc{Z}_{B,N-k}^{(m_1,m_2|n_1,n_2)}(q) \, ,  
\end{split}
\eeq
where
\begin{align}
&E_k^{(m_1,m_2|n_1,n_2)}(x=1,\bar{x}=1;y=1,\bar{y}=1;q) \nn \\ 
&~~~~~~~~~~~~~~~~~~~~~~= \sum\limits_{r=0}^k \sum\limits_{s=0}^r \sum\limits_{t=0}^{k-r} q^{k-t-s}~ C^{m_1}_s~ C^{m_2}_{r-s}~ C^{n_1+t-1}_t~ C^{n_2+k-r-t-1}_{k-r-t}\, ,
\end{align}
obtained by substituting $x=1,\bar{x}=1;y=1,\bar{y}=1$ in \Eq{super_el_q},
 and subsequently using Eqs. \eq{elval}, \eq{comval},   \eq{elpolq} and \eq{com_pol_q}.
\noi \section{\texorpdfstring{q-Deformed and restricted super Schur polynomials}{BorderSchur}}
\renewcommand{\theequation}{4.{\arabic{equation}}}
\setcounter{equation}{0}
\medskip

It has been found earlier that, super Schur polynomials 
associated with border strips 
play a key role in classifying the degenerate multiplets 
appearing in the spectra of $A_{N-1}$ type of \su{m|n} 
supersymmetric PF spin chains \cite{HB00}. For the purpose 
of doing a similar classification in the case of $BC_N$ type 
of PF spin chains, in this section
we shall prescribe a Jacobi-Trudi like formula
to define  a $q$-deformed version of the super Schur polynomials 
in terms of the $q$-deformed elementary supersymmetric  polynomials 
\eq{super_el_q}. 
Subsequently, by expanding such $q$-deformed super Schur polynomials 
as a power series of the parameter $q$, we shall obtain the so called   `restricted' super Schur polynomials which are independent 
of $q$. Moreover, we shall present some combinatorial forms for computing 
both of the $q$-deformed and the restricted super Schur polynomials.

%It is already known that for the case of $A_{N-1}$ type of \su{m|n} PF 
%spin chains \eq{a1}, the supersymmetric elementary polynomials can be 
%used to define the super Schur polynomials associated with border 
%strips \cite{HB00}. 
%A border strip is a connected super skew Young diagram which do not contain 
%any $2 \times 2$ square box. 
It may be noted that, border strips 
%or  skew Young diagrams 
represent a class of irreducible representations of the
$Y(gl(m))$ Yangian algebra 
($Y(gl(m|n))$ super Yangian algebra), 
which span the Fock spaces of \su{m} (\su{m|n}) 
HS and PF spin chains \cite{KKN97,HB00}.
There exists a one-to-one correspondence between these 
border strips and the 
motifs which we have mentioned earlier. 
For a 
spin chain with $N$ lattice sites, a border strip is denoted as 
$\langle k_1,...,k_r \rangle$, 
where $k_i$'s are some positive integers satisfying the relation  
$\sum\limits_{i=1}^r k_r = N$ (see Fig.~\ref{bstrip}), and a motif 
is denoted as   $(\de_1, \de_2,...,\de_{N-1})$, where $\de_i \in \{ 0,1 \}$.
The mapping from such a border strip to a motif may now be defined as 
\beq
\label{bo_mot}
\langle k_1,...,k_r  \rangle \Rightarrow (\underbrace{0,...,0}_{k_1-1}, 1, \underbrace{0,...,0}_{k_2-1}, 1, ...., 1, \underbrace{0,...,0}_{k_r-1} )\, .
\eeq
%with $E_{\langle k_1,...,k_r | l \rangle} = E_{(\de_1, \de_2,...,\de_{N-1}|)}$. 
The inverse mapping from a motif to a border strip can be obtained by reading a motif $(\delta_1,\delta_2,....,\delta_{N-1})$ from the left and adding a box under (resp. left) the box when $\delta_j=0$ (resp. $\delta_j=1$) is encountered.

\tikzset{
    level/.style = {
        thick
    },
    connect/.style = {
        thick
    },
    notice/.style = {
        draw,
        rectangle callout,
        callout relative pointer={#1}
    },
    label/.style = {
        text width=2cm
    }
}

\begin{figure}
\centering
\begin{tikzpicture}
\draw[level] (0,0) -- (0.7,0);
\draw[level] (-0.7,-3.5) -- (0.7,-3.5);
\draw[connect] (0,0) -- (0,-5.5);
\draw[connect] (0.7,0) -- (0.7,-3.5);
\draw[level] (0,-0.7) -- (0.7,-0.7);
\draw[level] (0,-1.4) -- (0.7,-1.4);
\draw[level] (-0.7,-2.8) -- (0.7,-2.8);
\draw[connect] (-0.7,-2.8) -- (-0.7,-7.5);
\draw[level] (-1.4,-4.8) -- (0,-4.8);
\draw[level] (-1.4,-5.5) -- (0,-5.5);
\draw[connect] (-1.4,-4.8) -- (-1.4,-9);
\draw[level] (-2.1,-6.8) -- (-0.7,-6.8);
\draw[level] (-2.1,-7.5) -- (-0.7,-7.5);
\draw[connect] (-2.1,-6.8) -- (-2.1,-9);
\node[level] at (0.35,-1.8) {$\cdot$};
\node[level] at (0.35,-2.1) {$\cdot$};
\node[level] at (0.35,-2.4) {$\cdot$};
\node[level] at (-0.35,-3.8) {$\cdot$};
\node[level] at (-0.35,-4.1) {$\cdot$};
\node[level] at (-0.35,-4.4) {$\cdot$};
\node[level] at (-1.05,-5.8) {$\cdot$};
\node[level] at (-1.05,-6.1) {$\cdot$};
\node[level] at (-1.05,-6.4) {$\cdot$};
\node[level] at (-1.75,-7.8) {$\cdot$};
\node[level] at (-1.75,-8.1) {$\cdot$};
\node[level] at (-1.75,-8.4) {$\cdot$};
\node[level] at (2,-1.75) {$k_1$};
\node[level] at (2,-0.5) {$\bigg \Uparrow$};
\node[level] at (2,-3) {$\bigg \Downarrow$};
\node[level] at (3.3,-4.15) {$k_2$};
\node[level] at (3.3,-3.3) {$\bigg \Uparrow$};
\node[level] at (3.3,-5) {$\bigg \Downarrow$};
\node[level] at (4.6,-6.15) {$k_3$};
\node[level] at (4.6,-5.3) {$\bigg \Uparrow$};
\node[level] at (4.6,-7) {$\bigg \Downarrow$};
\node[level] at (-4.5,-4) {\scalebox{1.2}{$\langle k_1,k_2,...,k_r \rangle =$}};
\end{tikzpicture}  
\caption{Shape of the boarder strip $\langle k_1,...,k_r \rangle$, 
which can be mapped to a motif through \Eq{bo_mot}.}
\label{bstrip}
\end{figure}
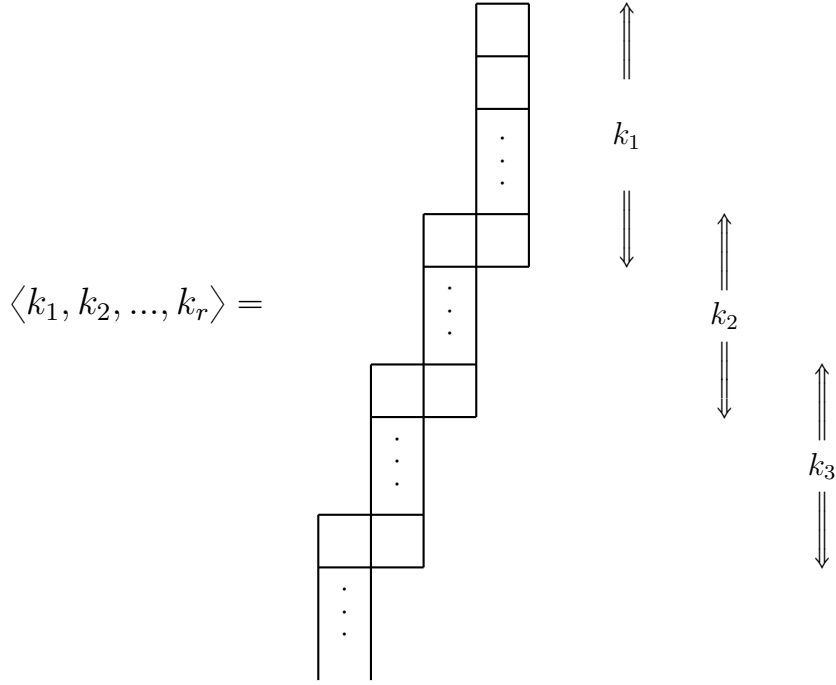
For the case of $A_{N-1}$ type of \su{m|n} PF spin chains, the super Schur polynomial associated with the border strip $\langle k_1,k_2,...,k_r \rangle$ is defined by using the following Jacobi-Trudi like determinant formula 
\cite{HB00}:
\beq
%S_{\langle k_1,k_2,....,k_r \rangle}^{(m|n)}(\mbf{x},\mbf{y})  =  
S_{\langle k_1,k_2,....,k_r \rangle}^{(m|n)}(x,\bar{x};y,\bar{y})  =  \begin{vmatrix}
E_{k_r} & E_{k_r+k_{r-1}} & \cdots & \cdots & E_{k_r+...+k_1}\\
1 & E_{k_{r-1}} & E_{k_{r-1}+k_{r-2}} & \cdots & E_{k_{r-1}+...+k_1}\\
0 & 1 & E_{k_{r-2}} & \cdots &  E_{k_{r-2}+...+k_1}\\
\vdots & \ddots & \ddots & \ddots & \vdots\\
0 & \cdots & 0 & 1 & E_{k_1}
\end{vmatrix} \, ,
\label{Schurdet}
\eeq
where $x\equiv x_1,x_2,\cdots,x_{m_1}$ and 
$\bar x\equiv \bar x_1, \bar x_2,\cdots, \bar x_{m_2}$ are two sets of bosonic
variables  (with total number  $m=m_1+m_2$), 
$y\equiv y_1,y_2,\cdots,y_{n_1}$ and 
$\bar y\equiv \bar y_1, \bar y_2,\cdots, \bar y_{n_2}$
are two sets of fermionic variables  (with total number  $n=n_1+n_2$), and 
the shorthand notation $E_k $  is used
for the elementary  supersymmetric  polynomial $ E_k^{(m|n)}(\mbf{x};\mbf{y}) $ (with $\mbf{x} \equiv x, \bar{x}$ and $\mbf{y} \equiv y, \bar{y}$)
defined in \eq{elsuA}.
%in \eq{elsuA}. 
In the limit $x=1,\bar{x}=1, y=1,\bar{y}=1$,  this super Schur polynomial reduces to
\beq
S_{\langle k_1,k_2,....,k_r \rangle}^{(m|n)}(x,\bar{x};y,\bar{y})|_{x=1,\bar{x}=1, y=1,\bar{y}=1 } =  \mc{N}^{(m|n)}_{ \langle k_1,k_2,...,k_r \rangle} \, ,  
\label{Schur_N}
\eeq
where  
$\mc{N}^{(m|n)}_{ \langle k_1,k_2,...,k_r \rangle} \in \Z$ gives  
the dimensionality of the irreducible representation of
$Y(gl(m|n))$ super Yangian algebra 
associated
with the border strip $\langle k_1,...,k_r \rangle$ or the corresponding motif. Hence, 
 $\mc{N}^{(m|n)}_{ \langle k_1,k_2,...,k_r \rangle}$'s also determine the dimensions
of the degenerate eigenspaces associated with the spectra of
the Yangian invariant  $A_{N-1}$ type of \su{m|n} PF spin chains.

 In analogy with the Jacobi-Trudi like determinant formula 
 \eq{Schurdet}, 
 let us define a novel $q$-dependent 
 super Schur polynomial associated with the border strip $\langle k_1,...,k_r 
 \rangle$ by using the $q$-deformed elementary supersymmetric  polynomials in \eq{super_el_q} as
\bea
&&S_{\langle k_1,k_2,....,k_r \rangle}^{(m_1,m_2|n_1,n_2)}(x,\bar{x};y,\bar{y};q) =  \nn \\
&& ~~~~~~~~~~~~~~~~~\begin{vmatrix}
E_{k_r}(q) & E_{k_r+k_{r-1}}(q) & \cdots & \cdots & E_{k_r+...+k_1}(q)\\
1 & E_{k_{r-1}}(q) & E_{k_{r-1}+k_{r-2}}(q) & \cdots & E_{k_{r-1}+...+k_1}(q)\\
0 & 1 & E_{k_{r-2}}(q) & \cdots &  E_{k_{r-2}+...+k_1}(q)\\
\vdots & \ddots & \ddots & \ddots & \vdots\\
0 & \cdots & 0 & 1 & E_{k_1}(q)
\end{vmatrix} \, ,
\label{Schurder}
\eea
where the notation $E_k(q) \equiv E_k^{(m_1,m_2|n_1,n_2)}(x,\bar{x};y,\bar{y};q) \, $  is used.
By putting $q=1$ in the above determinant and using Eqs. \eq{E_q_1} and \eq{Schurdet} consecutively, it can be readily seen that 
\beq
S_{\langle k_1,k_2,....,k_r \rangle}^{(m_1,m_2|n_1,n_2)}(x,\bar{x};y,\bar{y};q=1) = S_{\langle k_1,k_2,....,k_r \rangle}^{(m|n)}(x,\bar{x};y,\bar{y}).
\label{Sq1}
\eeq
Thus $S_{\langle k_1,k_2,....,k_r \rangle}^{(m_1,m_2|n_1,n_2)}(x,\bar{x};y,\bar{y};q)$ may be interpreted as a $q$-deformation of the super Schur polynomial $S_{\langle k_1,k_2,....,k_r \rangle}^{(m|n)}(x,\bar{x};y,\bar{y})$. 
Moreover, by using \Eq{x_qx_y_qy} along with \eq{Schurdet} 
and \eq{Schurder}, we obtain 
a relation like  
\beq
\label{Sxqxyqy}
S_{\langle k_1,k_2,....,k_r \rangle}^{(m_1,m_2|n_1,n_2)}(x,\bar{x};y,\bar{y};q) = S_{\langle k_1,k_2,....,k_r \rangle}^{(m|n)}(x,\bar{x};y,\bar{y})|_{\bar{x} \rightarrow q\bar{x}, \bar{y} \rightarrow q\bar{y}}.
\eeq
Since $S_{\langle k_1,k_2,....,k_r \rangle}^{(m_1,m_2|n_1,n_2)}(x,\bar{x};y,\bar{y})$ in 
\eq{Schurdet}
is a homogeneous polynomial of order $N$ \cite{HB00}, it follows from the above relation that
$S_{\langle k_1,k_2,....,k_r \rangle}^{(m_1,m_2|n_1,n_2)}(x,\bar{x};y,\bar{y};q)$ is also 
a homogeneous polynomial of order $N$ in the variables $x,\bar{x};y,\bar{y}$. 
Hence, \Eq{Sxqxyqy} may  be written in an alternative form like
\beq
S_{\langle k_1,k_2,....,k_r \rangle}^{(m_1,m_2|n_1,n_2)}(x,\bar{x};y,\bar{y};q) = q^N \cdot S_{\langle k_1,k_2,....,k_r \rangle}^{(m|n)}(x,\bar{x};y,\bar{y})|_{x \rightarrow q^{-1}x, y \rightarrow q^{-1}y}.
\label{Sxqxyqy2}
\eeq
From \Eq{Sxqxyqy} it is also evident that, 
$S_{\langle k_1,k_2,....,k_r \rangle}^{(m_1,m_2|n_1,n_2)}(x,\bar{x};y,\bar{y};q)$ is a  polynomial of order less than or equal to $N$ in the variable $q$. Consequently,
one can formally expand
$S_{\langle k_1,k_2,....,k_r \rangle}^{(m_1,m_2|n_1,n_2)}(x,\bar{x};y,\bar{y};q)$  in a power series of $q$ as
%Now, \Eq{comb_q} may be written as 
\beq
\label{Schur_l}
S_{\langle k_1,k_2,....,k_r \rangle}^{(m_1,m_2|n_1,n_2)}(x,\bar{x};y,\bar{y};q) = \sum\limits_{l=0}^N q^l \cdot S_{\langle k_1,k_2,....,k_r|l \rangle}^{(m_1,m_2|n_1,n_2)}(x,\bar{x};y,\bar{y})\, ,
\eeq
where, according to the terminology used by us,
the symbol $\langle k_1,k_2,....,k_r | l \rangle$ 
denotes a `branched' border strip (more precisely, the
$l$-th branch of the border strip $\langle k_1,k_2,....,k_r  \rangle$)
and 
$S_{\langle k_1,k_2,....,k_r|l \rangle}^{(m_1,m_2|n_1,n_2)}(x,\bar{x};y,\bar{y})$
denotes a `restricted' super Schur polynomial corresponding to the branched border strip $\langle k_1,k_2,....,k_r | l \rangle$. It will be shown shortly that the 
branched border strip $\langle k_1,k_2,....,k_r | l \rangle$ can be described
through a set of skew Young tableaux.
By putting $q=1$ in \eq{Schur_l} and using \eq{Sq1}, we find that the super Schur polynomial  is related to these restricted super Schur polynomials as
\beq
S_{\langle k_1,k_2,....,k_r \rangle}^{(m|n)}(x,\bar{x};y,\bar{y}) =  \sum\limits_{l=0}^N S_{\langle k_1,k_2,....,k_r|l \rangle}^{(m_1,m_2|n_1,n_2)}(x,\bar{x};y,\bar{y}) \, .
\label{sres}
\eeq
%Let us assume that
In the limit $x=1,\bar{x}=1, y=1,\bar{y}=1$,  the restricted super Schur polynomial 
yields
\beq
S_{\langle k_1,k_2,....,k_r|l \rangle}^{(m_1,m_2|n_1,n_2)}(x,\bar{x};y,\bar{y})|_{x=1,\bar{x}=1, y=1,\bar{y}=1 } =  \mc{N}^{(m_1,m_2|n_1,n_2)}_{ \langle k_1,k_2,...,k_r |l \rangle} \, ,  
\label{Schur_N_l}
\eeq
where  
$\mc{N}^{(m_1,m_2|n_1,n_2)}_{ \langle k_1,k_2,...,k_r | l \rangle}$'s are some non-negative integers which 
will be determined later. Inserting  $x = \bar x = y = \bar y = 1$ in 
\eq{sres}, and also using \eq{Schur_N} and  \eq{Schur_N_l}, we obtain 
\beq
\mc{N}^{(m|n)}_{\langle k_1,k_2,....,k_r \rangle} = \sum\limits_{l=0}^N \mc{N}^{(m_1,m_2|n_1,n_2)}_{\langle k_1,k_2,....,k_r|l \rangle}\, .
\label{split1}
\eeq
In analogy with the role played by 
$\mc{N}^{(m|n)}_{\langle k_1,k_2,....,k_r \rangle} $'s for the case of 
$A_{N-1}$ type of spin chains, we shall show in the next section that 
$\mc{N}^{(m_1,m_2|n_1,n_2)}_{\langle k_1,k_2,....,k_r|l \rangle}$'s  determine the dimensions
of the degenerate eigenspaces associated with the spectra of the 
$BC_N$ type of PF spin chains.
However, it seems to be  difficult    to explicitly 
compute $S_{\langle k_1,k_2,....,k_r|l \rangle}^{(m_1,m_2|n_1,n_2)}(x,\bar{x};y,\bar{y})$
and its limiting value $\mc{N}^{(m_1,m_2|n_1,n_2)}_{\langle k_1,k_2,....,k_r|l \rangle}$ 
by directly expanding $S_{\langle k_1,k_2,....,k_r \rangle}^{(m_1,m_2|n_1,n_2)}(x,\bar{x};y,\bar{y};q)$ in  \eq{Schurder}  as a power series of $q$. 

To bypass the above mentioned problem 
it may be noted that, apart from the determinant relation \eq{Schurdet},
the super Schur polynomial $S_{\langle k_1,k_2,....,k_r \rangle}^{(m|n)}(x,\bar{x};y,\bar{y})$ can also 
be expressed in the following combinatorial form \cite{BBHS07,BBH10}. To begin with, let us recall that a skew Young tableau associated with the border strip $\langle k_1,k_2,....,k_r \rangle$ is constructed by  
filling up this border strip with the numbers $1,2,....,m+n$, such that their arrangement obey two rules given by: 
\begin{enumerate}[label=(\roman*)]
    \item \label{rule1}{
    Entries in each row are increasing, allowing the repetition of elements in $\{i|i \in S_{+,+}^{(m_1)} \cup S_{+,-}^{(m_2)}\}$, but not permitting the repetition of elements in $\{i|i \in S_{-,+}^{(n_1)} \cup S_{-,-}^{(n_2)}\}$,
    }
    \item \label{rule2}{
    Entries in each column are increasing, allowing the repetition of elements in $\{i|i \in S_{-,+}^{(n_1)} \cup S_{-,-}^{(n_2)}\}$ but not permitting the repetition of elements in $\{i|i \in S_{+,+}^{(m_1)} \cup S_{+,-}^{(m_2)}\}$, 
    }
\end{enumerate}
where the sets $S_{+,+}^{(m_1)}$, $S_{+,-}^{(m_2)}$, $S_{-,+}^{(n_1)}$ and $S_{-,-}^{(n_2)}$ are defined in \eq{sets}. Let $\mathscr{G}$ be the set of all skew Young tableaux which are obtained by filling up the border strip $\langle k_1,k_2,....,k_r \rangle$ through the above mentioned rules. The combinatorial form of  $S_{\langle k_1,k_2,....,k_r \rangle}^{(m|n)}(x,\bar{x};y,\bar{y})$ may now be written as 
\beq
\label{comb}
S_{\langle k_1,k_2,....,k_r \rangle}^{(m|n)}(x,\bar{x};y,\bar{y}) = \sum\limits_{\mc{T} \in \, \msc{G}}  e^{wt(\mc{T})}\, ,
\eeq
where $wt(\mc{T})$ is the weight of the    
tableau $\mc{T}$ given by
\beq
wt(\mc{T}) = \sum\limits_{\al = 1}^{m+n} \al(\mc{T}) \cdot \ep_{\al} \, ,
\label{wt_T}
\eeq
$\al(\mc{T})$ denotes the multiplicity of the number $\al$ in the tableau $\mc{T}$, and the four sets of  variables  $x$, $\bar{x}$, $y$, $\bar{y}$ are defined as
\beq
\label{epsilon}
\begin{split}
&x_{\al} \equiv e^{\ep_{\al}}\, , ~~ \rm{if} ~~ \al \in S_{+,+}^{(m_1)}\, ,\\
&\bar{x}_{\al - m_1} \equiv e^{\ep_{\al}} \, ,~~ \rm{if} ~~ \al \in S_{+,-}^{(m_2)}\, ,\\
&y_{\al-(m_1+m_2)} \equiv e^{\ep_{\al}} \, ,~~ \rm{if} ~~ \al \in S_{-,+}^{(n_1)}\, ,\\
&\bar{y}_{\al-(m_1+m_2+n_1)} \equiv e^{\ep_{\al}} \, ,
~~ \rm{if} ~~ \al \in S_{-,-}^{(n_2)} \, . 
\end{split}
\eeq
Inserting $x=1,\bar{x}=1, y=1,\bar{y}=1$ in \eq{comb} and comparing it with \eq{Schur_N},
it is easy to see that $\mc{N}^{(m|n)}_{ \langle k_1,k_2,...,k_r \rangle}$ coincides with  
 the number of distinct tableaux within the set $\mc{G}$.  

At present, our aim is  to suitably modify \Eq{comb} 
to obtain the combinatorial form for both of the 
$q$-deformed and restricted
super Schur polynomials. Since,
due to \Eq{Sxqxyqy}, $\bar{x}$ and $\bar{y}$ in $S_{\langle k_1,k_2,....,k_r \rangle}^{(m|n)}(x,\bar{x};y,\bar{y})$ should be replaced by $q\bar{x}$ and $q\bar{y}$ respectively to obtain $S_{\langle k_1,k_2,....,k_r \rangle}^{(m_1,m_2|n_1,n_2)}(x,\bar{x};y,\bar{y};q)$, the  combinatorial form of 
the $q$-deformed super Schur polynomial can easily 
be obtained by modifying  \Eq{comb} as 
\beq
\label{comb_q}
S_{\langle k_1,k_2,....,k_r \rangle}^{(m_1,m_2|n_1,n_2)}(x,\bar{x};y,\bar{y};q) = \sum\limits_{\mc{T} \in \msc{G}} q^{\msc{F}(\mc{T})}~ e^{wt(\mc{T})}\, ,
\eeq
where $\msc{F}(\mc{T})$ for the tableau $\mc{T}$ is given by
\beq
\label{tableaux}
\msc{F}(\mc{T}) = \sum\limits_{\al = 1}^{m_1+m_2+n_1+n_2} \al(\mc{T}) \, f^{(m_1,m_2|n_1,n_2)}(\al)\, ,
\eeq
and $f^{(m_1,m_2|n_1,n_2)}(\al)$ is defined as
\bea
f^{(m_1,m_2|n_1,n_2)}(\al) & = 0, ~~ \rm{if} ~~ \al 
\in S_{+,+}^{(m_1)} \cup S_{-,+}^{(n_1)}\, , \nn \\
& = 1, ~~ \rm{if} ~~ \al \in S_{+,-}^{(m_2)} \cup S_{-,-}^{(n_2)} \,  . 
\label{fspin}
\eea
From \Eq{tableaux}, it is evident that $\msc{F}(\mc{T})$ is a non-negative integer satisfying the relation $\msc{F}(\mc{T}) \leq N$. Let $\msc{G}_l$ be the set of all tableaux  which are obtained by filling up the border strip $\langle k_1,k_2,....,k_r \rangle$ through the previously mentioned rules \ref{rule1} and \ref{rule2}, and for which $\msc{F}(\mc{T}) = l$, i.e.,
\beq 
\msc{G}_l = \{ \, \mc{T} \in \msc{G} \, | \, \msc{F}(\mc{T}) = l 
\, \} \, .
\label{setgl}
\eeq
Hence, the set $\msc{G}$ may be written as  
\beq
\label{G_union}
\msc{G} = \bigcup\limits_{0 \leq l \leq N} \msc{G}_l \, ,
\eeq
and \eq{comb_q} can be expressed in the form
\beq
\label{comb_q1}
S_{\langle k_1,k_2,....,k_r \rangle}^{(m_1,m_2|n_1,n_2)}(x,\bar{x};y,\bar{y};q) = 
\sum_{l=0}^N q^l  \sum\limits_{\mc{T} \in \msc{G}_l} e^{wt(\mc{T})}  \, .
\eeq
Comparing the above equation with \eq{Schur_l}, we finally
obtain the combinatorial form 
of the restricted super Schur polynomials as 
\beq
S_{\langle k_1,k_2,....,k_r|l \rangle}^{(m_1,m_2|n_1,n_2)}(x,\bar{x};y,\bar{y}) = \sum\limits_{\mc{T} \in \msc{G}_l} e^{wt(\mc{T})} \, .
\label{Schur_wt}
\eeq
%where $wt(\mc{T})$ defined in \eq{wt_T}
%and the sets of  variables  $x$, $\bar{x}$, $y$, $\bar{y}$ are %defined in \eq{epsilon}.
The above equation clearly shows that all information about the branched border strip $\langle k_1,k_2,....,k_r|l \rangle$ is essentially 
encoded within the set $\msc{G}_l$.  
Substituting $x=1,\bar{x}=1, y=1,\bar{y}=1$ in \eq{Schur_wt} and 
using \eq{Schur_N_l}, we find that 
\beq 
\mc{N}^{(m_1,m_2|n_1,n_2)}_{\langle k_1,k_2,....,k_r|l \rangle}
= \v \msc{G}_l  \v \, , 
\label{Schur_wt1}
\eeq
where $ \v \msc{G}_l \v$ denotes the number of elements 
in the set $\msc{G}_l$. 

For the purpose of explaining the above mentioned procedure of 
computing the $q$-deformed and the restricted super Schur polynomials through a particular example, let us assume that $m_1=1$, $m_2=1$, $n_1=0$, $n_2=2$  and consider the border strip $\langle 2,1 \rangle$ for $N=3$.
In this case, the sets in \Eq{sets} are given by:
$S^{(1)}_{+,+}=\{1\}$, $S^{(1)}_{+,-}=\{2\}$,
$S^{(0)}_{-,+}=\{~\}$ and $S^{(2)}_{-,-}=\{3,4\}$. Moreover, by using \Eq{fspin}, 
we obtain
$f^{(1,1|0,2)}(1)=0$, $f^{(1,1|0,2)}(2)=f^{(1,1|0,2)}(3)=f^{(1,1|0,2)}(4)=1$. 
Now, for each value of $l \in \{0,1,2,3 \}$, we can construct all possible
tableaux $\mc{T}$ following the rules \ref{rule1} and \ref{rule2} for which  $\msc{F}(\mc{T}) = l$. 
For $l=0$, there will be no possible tableau in this case, i.e., $\msc{G}_0=\{~\}$.
For $l=1$, the set of tableaux are given by :
%\ytableausetup{notabloids}
\[ \msc{G}_1  \! = \!  \left \{ \,
%\begin{align*}
%1
\begin{ytableau}
\none & 1  \\
 1 & 2 
\end{ytableau} \, , \,
%2
\begin{ytableau}
\none & 1  \\
 1 & 3 
\end{ytableau}\, , \,
%3
\begin{ytableau}
\none & 1  \\
 1 & 4 
\end{ytableau}   
%\end{align*} 
~\right \} \, ,
\]
for $l=2$,
\[ \msc{G}_2  \! = \!  \left \{ \,
%\begin{align*}
%4
\begin{ytableau}
\none & 1  \\
 2 & 2 
\end{ytableau} \, , 
%5
\begin{ytableau}
\none & 1  \\
 2 & 3 
\end{ytableau} \, , \, 
%6
\begin{ytableau}
\none & 1  \\
 2 & 4 
\end{ytableau} \, , \,
%7
\begin{ytableau}
\none & 1  \\
 3 & 4
\end{ytableau} \, , \,
%8
\begin{ytableau}
\none & 2  \\
 1 & 3 
\end{ytableau} \, , \, 
%9
\begin{ytableau}
\none & 2  \\
 1 & 4 
\end{ytableau} \, , \, 
%10
\begin{ytableau}
\none & 3  \\
 1 & 3 
\end{ytableau} \, , \, 
%11
\begin{ytableau}
\none & 3  \\
 1 & 4 
\end{ytableau} \, , \, 
%12
\begin{ytableau}
\none & 4  \\
 1 & 4 
\end{ytableau}   
%\end{align*}
\, \right\} ,  \] \, 
and for $l=3$,
%\begin{align*}
%13
\[ \msc{G}_3  \! = \!  \left \{ \,
\begin{ytableau}
\none & 2  \\
 2 & 3 
\end{ytableau}\, , \,
%14
\begin{ytableau}
\none & 2  \\
 2 & 4 
\end{ytableau}\, , \,
%15
\begin{ytableau}
\none & 2  \\
 3 & 4 
\end{ytableau}\, , \,
%16
\begin{ytableau}
\none & 3  \\
 2 & 3 
\end{ytableau}\, , \,
%17
\begin{ytableau}
\none & 3  \\
 2 & 4 
\end{ytableau}\, , \,
%18
\begin{ytableau}
\none & 4  \\
 2 & 4 
\end{ytableau}\, , \,
%19
\begin{ytableau}
\none & 3  \\
 3 & 4 
\end{ytableau}\, , \,
%20
\begin{ytableau}
\none & 4  \\
3 & 4 
\end{ytableau}
\, \right\} . \] \,    
%\end{align*}
\\
Using the combinatorial expression \eq{Schur_wt}, the corresponding restricted super Schur polynomials are obtained as
%\begin{subequations}
\bea
\label{rssp}
&&S_{\langle 2,1|0 \rangle}^{(1,1|0,2)}(x_1,\bar{x}_1;\bar{y}_1,\bar{y}_2) = 0\, , \nn \\
&&S_{\langle 2,1|1 \rangle}^{(1,1|0,2)}(x_1,\bar{x}_1;\bar{y}_1,\bar{y}_2) = x_1^2(\bar{x}_1 + \bar{y}_1 + \bar{y}_2)\, ,
\nn \\
&&S_{\langle 2,1|2 \rangle}^{(1,1|0,2)}(x_1,\bar{x}_1;\bar{y}_1,\bar{y}_2) = x_1(\bar{x}_1 + \bar{y}_1 + \bar{y}_2)^2 \, ,
\nn \\
&&S_{\langle 2,1|3 \rangle}^{(1,1|0,2)}(x_1,\bar{x}_1;\bar{y}_1,\bar{y}_2) = (\bar{y}_1 + \bar{y}_2) (\bar{x}^2_1 + \bar{x}_1\bar{y}_1 + \bar{x}_1\bar{y}_2 + \bar{y}_1\bar{y}_2 ) \, .
\eea
%\end{subequations}
Inserting  $x_1 =\bar{x}_1=\bar{y}_1=\bar{y}_2=1$
in the above equation or directly using \Eq{Schur_wt1}, we 
obtain 
\beq 
\mc{N}_{\langle 2,1|0 \rangle}^{(1,1|0,2)}=0 \, ,~~ 
\mc{N}_{\langle 2,1|1 \rangle}^{(1,1|0,2)}=3 \, ,~~
\mc{N}_{\langle 2,1|2 \rangle}^{(1,1|0,2)}=9 \, ,~~
\mc{N}_{\langle 2,1|3 \rangle}^{(1,1|0,2)}=8 \, ,
\eeq
Moreover,  using Eqs. \eq{rssp} and \eq{Schur_l}, we can write the corresponding $q$-deformed super Schur polynomial as
\beq
\label{example}
\begin{split}
S_{\langle 2,1 \rangle}^{(1,1|0,2)}(x_1,\bar{x}_1;\bar{y}_1,\bar{y}_2;q) = q \cdot x_1^2(\bar{x}_1 + \bar{y}_1 + \bar{y}_2) + q^2 \cdot x_1(\bar{x}_1 + \bar{y}_1 + \bar{y}_2)^2 \\ + q^3 \cdot (\bar{y}_1 + \bar{y}_2) (\bar{x}^2_1 + \bar{x}_1\bar{y}_1 + \bar{x}_1\bar{y}_2 + \bar{y}_1\bar{y}_2 ).    
\end{split}
\eeq
%The same expression for this  $q$-deformed super Schur polynomial %can also be obtained from the  determinant relation  \eq{Schurder}.
In the next section, we shall explain how this type of    
 $q$-deformed and restricted super Schur polynomials 
 play an important role in classifying
 the degenerate multiplets  within the spectra of
 the $BC_N$ type of  PF spin chains.
%In the next section, we shall explain how one can map the branched border strips corresponding to the restricted Schur polynomials into branched motifs.

\noi \section{\texorpdfstring{Spectra of the $BC_N$ type of  ferromagnetic PF spin chains }{bordermotifs}}
\renewcommand{\theequation}{5.{\arabic{equation}}}
\setcounter{equation}{0}
\medskip
It has been found earlier that the $A_{N-1}$ type of homogeneous RS  polynomials can be expressed through some suitable linear combinations of the Schur polynomials associated with the border strips
\cite{KKN97}.  Similarly, the $A_{N-1}$ type of homogeneous SRS polynomials \eq{SRSA} can be expressed through super Schur polynomials
\eq{Schurdet}  as \cite{HB00}
\beq
\mbb{H}_{A,N}^{(m|n)}(\mbf{x},\mbf{y};q)=  \sum\limits_{\Vec{k} \in \mc{P}_N} q^{\frac{N(N-1)}{2} - \sum\limits_{i=1}^{r-1} K_i} \cdot \,
S_{\langle k_1,k_2,....,k_r \rangle}^{(m|n)}(x,\bar{x};y,\bar{y}) ,
\label{RS_0}
\eeq
where $\mbf{x}\equiv x, \bar{x}$, $\mbf{y}\equiv y, \bar{y}$, 
$\mc{P}_N$ denotes the set of all ordered partitions of $N$, $\Vec{k} \equiv \{ k_1,k_2,...,k_r \}$ (where $r$ is an integer taking value from $1$ to $N$) is an element of $\mc{P}_N$ and $K_i$'s are partial sums 
%(corresponding to some given $\Vec{k}$) 
given by $K_i = \sum\limits_{j=1}^i k_j$. 
Substituting $x=\bar{x}=y=\bar{y}=1$ in \eq{RS_0} and using \eq{Schur_N}, 
one obtains 
\beq
\mc{Z}_{A,N}^{(m|n)}(q) = 
\sum\limits_{\Vec{k} \in \mc{P}_N} q^{\frac{N(N-1)}{2} - \sum\limits_{i=1}^{r-1} K_i} 
\cdot \mc{N}^{(m|n)}_{\langle k_1,...,k_r  \rangle}\, ,
\label{rest_par0}
\eeq
where $\mc{Z}_{A,N}^{(m|n)}(q) $
%\equiv \mbb{H}_{A,N}^{(m|n)}
%(\mbf{x},\mbf{y};q)|_{\mbf{x}=1,\mbf{y}=1} 
denotes the partition function of the 
$A_{N-1}$ type of supersymmetric PF spin chain \eq{a1}. From the structure
of $\mc{Z}_{A,N}^{(m|n)}(q)$ in \eq{rest_par0} it is clear that, the exponent of 
$q$ yields
the energy level  for the spin chain \eq{a1} associated with the  border strip $\langle k_1,...,k_r  \rangle$ as 
\beq
\label{en_border0}
\mc{E}^{(m|n)}_{\langle k_1,...,k_r  \rangle} = \frac{N(N-1)}{2} - \sum\limits_{i=1}^{r-1} K_i \, ,
\eeq
and the degeneracy of this energy level is given by $\mc{N}^{(m|n)}_{\langle k_1,...,k_r  \rangle}$. 
Using the mapping \eq{bo_mot} between the border strips and the motifs, 
it is easy to show that
\beq
\label{par_motif}
\sum\limits_{i=1}^{r-1} K_i = \sum\limits_{j=1}^{N-1} j\de_j.
\eeq
Hence, by substituting \eq{par_motif} into \eq{en_border0}, one can 
write down all energy levels  of the supersymmetric PF spin chain \eq{a1}
in terms of motifs as
\beq
\label{en_motif0}
\mc{E}^{(m|n)}_{(\de_1, \de_2,...,\de_{N-1})} \equiv \mc{E}^{(m|n)}_{\langle k_1,...,k_r  \rangle} = \sum\limits_{j=1}^{N-1} j(1-\de_j) 
\eeq
with degeneracy factor 
$\mc{N}^{(m|n)}_ {(\de_1, \de_2,...,\de_{N-1})} 
\equiv \mc{N}^{(m|n)}_{\langle k_1,...,k_r \rangle}$.

At present, our aim is to express the $BC_N$ type of SRS polynomials 
\eq{SRS0} through the   
 $q$-deformed and the restricted super Schur polynomials which have been defined 
 in Sec.~4 and subsequently use such expression for classifying
 the degenerate multiplets  within the spectra of
 the $BC_N$ type of  ferromagnetic PF spin chains \eq{b8} with SAPSRO. 
 For this purpose, in analogy with 
 $\mbb{H}_{A,N}^{(m|n)}(\mbf{x},\mbf{y};q)$ appearing in \Eq{RS_0}, let us define a polynomial $\mc{F}^{(m_1,m_2|n_1,n_2)}_N(x,\bar x;y,\bar y;q)$ for $N>0$ as 
\beq
\label{RSnew0}
\mc{F}_{N}^{(m_1,m_2|n_1,n_2)}(x,\bar x;y,\bar y;q) = \sum\limits_{\Vec{k} \in \mc{P}_N} q^{N(N-1) - 2\sum\limits_{i=1}^{r-1} K_i} \cdot S_{\langle k_1,...,k_r \rangle}^{(m_1,m_2|n_1,n_2)}(x,\bar x;y,\bar y;q)\, ,
\eeq
%where $\mc{P}_N$ denotes the set of all ordered partitions of $N$, $\Vec{k}  %\equiv \{ k_1,k_2,...,k_r \}$ (where $r$ is an integer with value from $1$ to %$N$  ) is an element of $\mc{P}_N$ and $K_i$'s are partial sums 
%(corresponding to some given $\Vec{k}$) 
%given by $K_i = \sum\limits_{j=1}^i k_j$.   
and also assume that 
$ \mc{F}_{0}^{(m_1,m_2|n_1,n_2)}(x,\bar x;y,\bar y;q) =1$. 
Let us now evaluate the polynomial $\mc{F}^{(m_1,m_2|n_1,n_2)}_N(x,\bar x;y,\bar y;q)$ 
%given by \eq{RSnew0}
for the cases $N=1,2$ and $3$, by using shorthand notations like  
\[
\mc{F}_N(q) \equiv 
\mc{F}^{(m_1,m_2|n_1,n_2)}_N(x,\bar x;y,\bar y;q) \, , ~~
S_{\langle k_1,...,k_r \rangle} (q) \equiv  S_{\langle k_1,...,k_r \rangle}^{(m_1,m_2|n_1,n_2)} (x,\bar{x};y,\bar{y};q) \, ,
\]
for sake of brevity.  
For the case $N=1$, Eqs.~\eq{RSnew0} and \eq{Schurder} imply  that
\bea
\label{RSn1}
\mc{F}_1(q) =S_{\langle 1 \rangle}(q)= E_1(q)\, .
\eea
Similarly, for the cases $N=2$ and $N=3$, we obtain  
%\[
%\mc{F}_2(q) = 
%q^2 \cdot S_{\langle 2 \rangle} (q) + S_{\langle 1,1 \rangle} (q)\, ,
%\eea
%where . Now, using \eq{Schurder}, we have
\begin{align}
\mc{F}_2(q) &= q^2 \cdot S_{\langle 2 \rangle} (q) +
S_{\langle 1,1 \rangle} (q) \nn \\
& = q^2 \cdot E_2 (q) + 
   \begin{vmatrix}
   E_1(q) & E_2(q)\\
   1 & E_1(q)
   \end{vmatrix} \nn \\
   & = E_1^2(q) + (q^2-1)E_2(q) \, ,
   \label{RSn2}
   \end{align}
%Similarly, for $N=3$,
  and 
  \begin{align}
    \mc{F}_3(q) &= q^6 \cdot S_{\langle 3 \rangle}(q) + q^4 \cdot S_{\langle 1,2 \rangle}(q) + q^2 \cdot S_{\langle 2,1 \rangle}(q) + S_{\langle 1,1,1 \rangle}(q) \notag \\
    \nn \\
   & = q^6 \cdot E_3(q)  + q^4 \cdot 
   \begin{vmatrix}
   E_2(q) & E_3(q) \\
   1 & E_1(q)
   \end{vmatrix} + q^2 \cdot 
   \begin{vmatrix}
   E_1(q) & E_3(q)\\
   1 & E_2(q)
   \end{vmatrix} 
   \nn \\
   \nn \\
   &~~~~~~~~~~~~~~~~~~~~~~~~~~~~~~~~~~~~~~~~~~~~~~+
   \begin{vmatrix}
   E_1(q) & E_2(q) & E_3(q)\\
   1 & E_1(q) & E_2(q)\\
   0 & 1 & E_1(q)
   \end{vmatrix}
   \nn \\
   \nn \\
   &= E_1^3(q)+(q^2-1)(q^2+2)E_1(q)E_2(q)+(1-q^2)(1-q^4)E_3(q).
   \label{RSn3}
\end{align}
It can be readily seen that the right hand sides of Eqs.~\eq{RSn1}, \eq{RSn2} and \eq{RSn3} exactly match with  those of  Eqs.~\eq{RS1}, \eq{RS2} and \eq{RS3} respectively. This result strongly suggests that the equality  
\beq
\label{rsequ}
\mbb{H}^{(m_1,m_2|n_1,n_2)}_N(x,\bar x;y,\bar y;q) = \mc{F}^{(m_1,m_2|n_1,n_2)}_N(x,\bar x;y,\bar y;q) \, , 
\eeq
would hold for arbitrary values of $N$. In the following, we shall give a proof of this statement
by using a procedure similar to what has been described in Appendix A of Ref.~\cite{KKN97} for the 
case of $A_{N-1}$ type of PF spin chain. 

Expanding the determinant formula \eq{Schurder} along its first row,  we obtain
a recursion relation  for the $q$-deformed super Schur polynomials as 
\beq
\label{Schurrecur}
S_{\langle k_1,k_2,....,k_r \rangle}^{(m_1,m_2|n_1,n_2)}(x,\bar{x};y,\bar{y};q) = \sum\limits_{r=1}^s (-1)^{s+1}\, E_{k_r + k_{r-1} + \cdots + k_{r-s+1}}(q)\, S_{\langle k_1,k_2,....,k_{r-s} \rangle}^{(m_1,m_2|n_1,n_2)}(x,\bar{x};y,\bar{y};q). 
\eeq
Substituting the above recursion relation into \Eq{RSnew0}, we can  rewrite the polynomial $\mc{F}^{(m_1,m_2|n_1,n_2)}_N(x,\bar x;y,\bar y;q)$ in the following form for any $N\geq 1$:
\beq
\label{core}
\mc{F}^{(m_1,m_2|n_1,n_2)}_N(x,\bar x;y,\bar y;q)=\sum\limits_{k=1}^N A_{N,k}(q^2)\, E_k(q)\, \mc{F}^{(m_1,m_2|n_1,n_2)}_{N-k}(x,\bar x;y,\bar y;q)\, ,
\eeq
where 
\begin{equation}
A_{N,k}(q^2) = \sum\limits_{j=1}^k \sum\limits_{l_1+...+l_j=k} (-1)^{j+1} (q^2)^{C_{N,k}(l_1,...,l_j)} \, ,
\label{core0}
\end{equation}
with 
\begin{align*}
    C_{N,k}(l_1,...,l_j) & = \frac{1}{2}N(N+1)-\frac{1}{2}(N-k)(N-k+1) - \sum\limits_{i=1}^j (l_1+...+l_i+N-k) \\
    & = N(k-j)-\frac{1}{2}k(k+1)+\sum\limits_{i=1}^j i l_i.
\end{align*}
It may be noted that exactly this form of $C_{N,k}(l_1,...,l_j)$ has appeared 
earlier in Eq.~(A.6) of Ref.~\cite{KKN97}. Hence, by using the 
method of latter reference, we compute the function $A_{N,k}(q^2)$ in 
\eq{core0} as 
\begin{equation*}
A_{N,k}(q^2) = (-1)^{k+1} \frac{(q^2)_{N-1}}{(q^2)_{N-k}}.
\end{equation*}
Substituting the above expression of $ A_{N,k}(q^2)$ 
into \Eq{core}, we find that the polynomial $\mc{F}^{(m_1,m_2|n_1,n_2)}_N(x,\bar x;y,\bar y;q)$ satisfies the recursion relation
\beq
\mc{F}^{(m_1,m_2|n_1,n_2)}_N(x,\bar x;y,\bar y;q)=\sum\limits_{k=1}^N (-1)^{k+1} \frac{(q^2)_{N-1}}{(q^2)_{N-k}}\, E_k(q)\, \mc{F}^{(m_1,m_2|n_1,n_2)}_{N-k}(x,\bar x;y,\bar y;q) \, . 
\eeq
It may be noted that the form of above recursion relation 
coincides with that of \eq{RS_recursion} satisfied by the $BC_N$ type of SRS polynomials. Since the $BC_N$ type of SRS polynomials are uniquely determined
by the recursion relation  \eq{RS_recursion} and the initial condition
 $\mbb{H}_{B,0}^{(m_1,m_2|n_1,n_2)}(x,\bar x;y,\bar y;q)=1$, the equality
 \eq{rsequ} is proved for arbitrary values of $N$. 

Combining Eqs.~\eq{rsequ} and \eq{RSnew0}, we obtain the desired expression
 for  $BC_N$ type of SRS polynomials in terms of linear combinations of 
 $q$-deformed super Schur polynomials 
as  
\begin{align}
\mbb{H}^{(m_1,m_2|n_1,n_2)}_N(x,\bar x;y,\bar y;q) 
%& = \mc{F}^{(m_1,m_2|n_1,n_2)}_N(x,\bar x;y,\bar y;q) \nn  \\
& = \sum\limits_{\Vec{k} \in \mc{P}_N} q^{N(N-1) - 2\sum\limits_{i=1}^{r-1} K_i} \cdot S_{\langle k_1,...,k_r \rangle}^{(m_1,m_2|n_1,n_2)}(x,\bar x;y,\bar y;q).  
\label{RS_new}
\end{align}
Substituting the expansion of  $S_{\langle k_1,...,k_r \rangle}^{(m_1,m_2|n_1,n_2)}(x,\bar x;y,\bar y;q) $ in \eq{Schur_l} to
\eq{RS_new},  we further express
 the  $BC_N$ type of SRS polynomials through restricted 
super Schur polynomials as  
\beq
    \label{RS_l}
    \mbb{H}_{B,N}^{(m_1,m_2|n_1,n_2)}(x,\bar x,y,\bar y;q) = \sum\limits_{\Vec{k} \in \mc{P}_N} \sum\limits_{l=0}^N q^{N(N-1) - 2\sum\limits_{i=1}^{r-1} K_i + l} \cdot S^{(m_1,m_2|n_1,n_2)}_{\langle k_1,...,k_r | l \rangle}(x,\bar x,y,\bar y).
\eeq
Inserting $x=\bar{x}=y=\bar{y}=1$ in \Eq{RS_l} and subsequently using Eqs.~\eq{SRSpa} as well as \eq{Schur_N_l}, we find that the partition function of the $BC_N$ type of
PF spin chain \eq{b8} can be written as 
\beq
\mc{Z}_{B,N}^{(m_1,m_2|n_1,n_2)}(q) = \sum\limits_{\Vec{k} \in \mc{P}_N} \sum\limits_{l=0}^N q^{N(N-1) - 2\sum\limits_{i=1}^{r-1} K_i + l} \cdot \mc{N}^{(m_1,m_2|n_1,n_2)}_{\langle k_1,...,k_r | l \rangle} \, .
\label{rest_par}
\eeq
From the structure of $\mc{Z}_{B,N}^{(m_1,m_2|n_1,n_2)}(q)$ in 
\eq{rest_par} it is evident that, the exponent of $q$ yields
the energy level of the spin chain \eq{b8} corresponding 
to the branched border strip $\langle k_1,...,k_r | l \rangle$ as
\beq
\label{en_border}
\mc{E}^{(m_1,m_2|n_1,n_2)}_{\langle k_1,...,k_r | l \rangle} =
N(N-1) - 2\sum\limits_{i=1}^{r-1} K_i + l \, ,
\eeq
and the degeneracy of this energy level is given by $\mc{N}^{(m_1,m_2|n_1,n_2)}_{\langle k_1,...,k_r | l \rangle}$. In analogy with \Eq{bo_mot},
we may now construct a one-to-one mapping 
between branched border strips like $\langle k_1,...,k_r | l \rangle$ and  branched motifs like $(\de_1, \de_2,...,\de_{N-1}|l)$ as 
\beq
\label{bo_mot1}
\langle k_1,...,k_r |l \rangle \Rightarrow \left(\underbrace{0,...,0}_{k_1-1}, 1, \underbrace{0,...,0}_{k_2-1}, 1, ...., 1, \underbrace{0,...,0}_{k_r-1}|l \right)\, .
\eeq
Hence, by inserting \eq{par_motif} into \eq{en_border},
%and using 
%the notation $E_{(\de_1, \de_2,...,\de_{N-1}|)}  \equiv 
%E_{\langle k_1,...,k_r | l \rangle} $, 
we finally 
obtain the energy level  of the $BC_N$ type of PF spin chain with 
SAPSRO \eq{b8}  associated with the branched motif
$(\de_1, \de_2,...,\de_{N-1}|l)$ as 
\beq
\label{en_motif}
\mc{E}^{(m_1,m_2|n_1,n_2)}_{(\de_1, \de_2,...,\de_{N-1}|l)}\equiv 
\mc{E}^{(m_1,m_2|n_1,n_2)}_{\langle k_1,...,k_r | l \rangle} =
%= N(N-1) - 2\sum\limits_{j=1}^{N-1} j\de_j + l 
2\sum\limits_{j=1}^{N-1} j(1-\de_j) + l \, ,
\eeq
 with degeneracy factor given by 
$\mc{N}^{(m_1,m_2|n_1,n_2)}_ {(\de_1, \de_2,...,\de_{N-1}|l)} 
\equiv
\mc{N}^{(m_1,m_2|n_1,n_2)}_{\langle k_1,...,k_r | l \rangle},$
which can be computed by using \Eq{Schur_wt1}. 
One may observe that, 
the dimension of the Hilbert space associated with the spin chain  \eq{b8}
is given by $d_1=(m+n)^N$ and the highest possible number of  
distinct energy levels of the form \eq{en_motif} is given by $d_2=2^{N-1} (N+1)$. Since $d_1$ and $d_2$ satisfy the relation
\begin{equation*}
    \ln{\left(\frac{d_2}{d_1}\right)} =
\ln{\left(N+1\right)} - N\ln{\left(\frac{m+n}{2}\right)} - \ln{2}\,
\, ,
\end{equation*}
assuming $m+n>2$ and taking $N \rightarrow \infty$ limit   
we find that
$\ln{\left(\frac{d_2}{d_1}\right)} = \mc{O}\left(\ln{N}\right)-\mc{O}\left(N\right) \rightarrow -\infty$, i.e., $\frac{d_2}{d_1} \rightarrow 0$. This result clearly indicates that, in general,  
%in spite of having a broken $Y(gl(m|n))$ super Yangian symmetry, 
the energy levels  of $BC_N$ type of PF spin chain 
are highly degenerate for large values of $N$.

It should be noted that, if  the degeneracy factor
$\mc{N}^{(m_1,m_2|n_1,n_2)}_{\langle k_1,...,k_r | l \rangle}$
becomes zero for some choice of the discrete parameters $m_1, \, m_2, 
\, n_1, \, n_2$, and branched motif $(\de_1, \de_2,...,\de_{N-1}|l)$, 
then the energy level $\mc{E}^{(m_1,m_2|n_1,n_2)}_{(\de_1, \de_2,...,\de_{N-1}|l)}$ 
in \eq{en_motif} would be absent from the spectrum of the 
corresponding spin chain. For example, in the case of  $BC_N$ type 
of bosonic PF spin chain with $m_1+m_2=m\geq 2$
and $n_1=n_2=0$, rule \ref{rule1} of Sec.~4 implies that at most $m$ boxes
can placed in any column of a border strip. As a result, branched border strips like $\langle k_1,...,k_r|l \rangle$ with 
$k_i \leq m$ can only give nontrivial values of $\mc{N}^{(m_1,m_2|n_1,n_2)}_{\langle k_1,...,k_r | l \rangle}$. 
Hence, by using the mapping \eq{bo_mot1}, we find that 
branched motifs like $(\de_1, \de_2,...,\de_{N-1}|l)$ 
with more than $(m-1)$ consecutive $\de_i=0$ do not appear in the spectrum of this spin chain. Similarly, by using the rule \ref{rule2} of Sec.~4,  one can show that 
branched motifs like $(\de_1, \de_2,...,\de_{N-1}|l)$ with more than $(n-1)$ consecutive $\de_i=1$ do not appear in the spectrum of the $BC_N$ type of 
fermionic PF spin chain with 
$m_1=m_2=0$ and  $n_1+n_2=n\geq 2$. Similar type of `selection
rules,' giving restrictions on possible values of $\delta_i$ within the motif
 $(\de_1, \de_2,...,\de_{N-1})$, 
have been found earlier in the context of $A_{N-1}$ type of non-supersymmetric HS and PF spin chains \cite {Ha93, Hi95}.

However it is worth noting that, apart from the above mentioned restrictions on 
possible values of $\delta_i$, some restriction on allowed values of $l$
may also be present for the case of branched motif 
$(\de_1, \de_2,...,\de_{N-1}|l)$. 
Such restriction on possible values of $l$ leads to a new type of selection rule for the case of some  $BC_N$ type of PF spin chains. 
To demonstrate the existence 
of the latter type of selection rule through a simple example, 
let us consider a  $BC_N$ type of PF spin chain \eq{b8}
with discrete parameters given by
$m_1=m$,  $m_2=0$, $n_1=n$, $n_2=0$, where $m+n\geq 2$. 
In this case, \Eq{SA_parity} implies that $f(s_i)=0$ for all possible value of the local spin $s_i$. 
As a result, $P^{(m,0|n,0)}_i$ defined in \eq{b7} becomes an identity operator and 
$\widetilde{P}_{ij}^{(m,0|n,0)}$ reduces to the spin permutation operator $P_{ij}^{(m|n)}$.
Therefore, in this special case,
%for $m_2=0$ and $n_2=0$, 
the Hamiltonian \eq{b8} of the $BC_N$ type of PF spin chain can be expressed in a simple form like 
\beq
\mc{H}^{(m,0|n,0)}_N
=\sum_{i\neq j} \frac{y_i+y_j}{(y_i-y_j)^2}  (1- P_{ij}^{(m|n)}) \, ,
\label{ys}
\eeq
where $y_j$'s are the roots of the the generalized Laguerre polynomial 
$L_N^{\beta -1}$ as mentioned earlier.
Similarly it can be shown that, in another special case with discrete parameters given by $m_1=0$,  $m_2=m$, $n_1=0$, $n_2=n$,  the corresponding Hamiltonian $\mc{H}^{(0,m|0,n)}_N$  would coincide with  
$\mc{H}^{(m,0|n,0)}_N$ in \eq{ys} up to an insignificant additive constant.

For the purpose of finding out the allowed branched motifs associated with  Hamiltonian $\mc{H}^{(m,0|n,0)}_N$ in \eq{ys}, we use \Eq{fspin} to  find that $f^{(m,0|n,0)}(\al) =0$ 
for all  possible value of $\al$.  Hence, Eqs.~\eq{tableaux}, \eq{setgl} and 
\eq{Schur_wt1} imply that 
$\mc{F}(\mc{T})=0$ for any $\mc{T} \in \mc{G}$, $\mc{G}_l=\{ \,\}$ for $l>0$,
and $\mc{N}^{(m,0|n,0)}_{\langle k_1,...,k_r | l \rangle}=0$ for $l>0$. 
As a consequence, the spectrum of $\mc{H}^{(m,0|n,0)}_N$ in \eq{ys}
obeys a selection rule which 
forbids the occurrence  of branched motifs like  $(\de_1, \de_2,...,\de_{N-1}|l)$ 
 for $l > 0$. Therefore, by putting $l=0$ in \eq{en_motif}, we find that the allowed energy levels of the Hamiltonian  $\mc{H}^{(m,0|n,0)}_N$ can
 in general be written in the form
\beq
\label{en_motif_v}
\mc{E}^{(m,0|n,0)}_{(\de_1, \de_2,...,\de_{N-1}|0)}=
 2\sum\limits_{j=1}^{N-1} j(1-\de_j) \, .
\eeq
 Using the  rules \ref{rule1} 
  and \ref{rule2} of Sec.~4, it is easy to see that  
  $\de_i$'s in the above equation can be chosen 
  without any restriction when both  
  $m$ and $n$ take nonzero values. 
 On the other hand, as
 we have already mentioned earlier, more than ($m-1$) number of 
 ($(n-1)$ number of) consecutive $\de_i=0$ ($\de_i=1$) 
  would not appear in the above equation
  in the special case  $n=0$ ($m=0$).
Comparing \Eq{en_motif_v} with \Eq{en_motif0}, we interestingly find that 
\beq
\label{eqeng}
\mc{E}^{(m,0|n,0)}_{(\de_1, \de_2,...,\de_{N-1}|0)} = 2 \, 
\mc{E}^{(m|n)}_{(\de_1, \de_2,...,\de_{N-1})} \, , 
\eeq
and, by inserting $\mc{N}^{(m,0|n,0)}_{\langle k_1,...,k_r | l \rangle}=0$ for $l>0$ in \Eq{split1}, we also get  
\beq
\label{eqdeg}
\mc{N}^{(m,0|n,0)}_{\langle k_1,k_2,....,k_r|0 \rangle} =
 \mc{N}^{(m|n)}_{\langle k_1,k_2,....,k_r \rangle} \, .
\eeq
Eqs.~\eq{eqeng} and \eq{eqdeg} clearly show that, except for an overall  scale factor of two, the spectrum of  the  
$BC_N$ type of PF chain with Hamiltonian  
$\mc{H}^{(m,0|n,0)}_N$ in  
\eq{ys} coincides with that of the $A_{N-1}$ type of supersymmetric PF chain with Hamiltonian $\mc{H}_N^{(m|n)}$ in 
\eq{a1}. Thus the Hamiltonians $\mc{H}^{(m,0|n,0)}_N$ and 
$2\mc{H}_N^{(m|n)}$ must be connected through a unitary transformation.
It is well known that the latter Hamiltonian exhibits $Y(gl(m))$ Yangian 
symmetry in the special case  $n=0$, $Y(gl(n))$ Yangian 
symmetry in the special case  $m=0$, and  $Y(gl(m|n))$ super Yangian symmetry when both $m$ and $n$ take nonzero values. Hence, 
Eqs.~\eq{eqeng} and \eq{eqdeg} imply that  
the Hamiltonian $\mc{H}^{(m,0|n,0)}_N$  (and the related 
Hamiltonian $\mc{H}^{(0,m|0,n)}_N$) would also exhibit $Y(gl(m))$ Yangian 
symmetry in the special case $n=0$, $Y(gl(n))$ Yangian 
symmetry in the special case  $m=0$, 
and $Y(gl(m|n))$ super Yangian symmetry when 
both $m$ and $n$ take nonzero values.

Let us now consider the class of Hamiltonians of 
the $BC_N$ type of PF spin chains \eq{b8},
which can not be expressed in the above mentioned
simple forms $\mc{H}^{(m,0|n,0)}_N$
and $\mc{H}^{(0,m|0,n)}_N$. 
It is interesting to observe that, 
%for any possible values of the discrete parameters  
%$m_1, \, m_2, \, n_1, \, n_2$, 
 \Eq{split1}  establishes a connection between  
the spectrum of  a $BC_N$ type of PF chain with Hamiltonian  
$\mc{H}^{(m_1,m_2|n_1,n_2)}_N$ and that of the $A_{N-1}$ type of supersymmetric PF chain with Hamiltonian $\mc{H}_N^{(m|n)}$  in 
\eq{a1}, where $m=m_1+m_2$ and $n=n_1+n_2$. More precisely, \Eq{split1}
shows that the degeneracy factor of the energy level (of the later spin chain) associated with the motif 
$(\de_1, \de_2,...,\de_{N-1})$ is exactly same as
the sum of 
degeneracy factors of all possible energy levels (of the former spin chain)   
associated with branched motifs of the form $(\de_1, \de_2,...,\de_{N-1}|l)$
over the variable $l$. 
%within the spectrum of the supersymmetric PF spin chain \eq{a1}. 
This type of splitting of energy levels, between two quantum systems
occupying the same Hilbert space, can only appear when 
an irreducible representation associated with the bigger symmetry algebra of
one quantum system is expressed as a direct sum of several irreducible representations 
associated with the smaller symmetry subalgebra of the other quantum system.
 Hence we conclude that, the symmetry algebra 
of the presently considered class of  
$BC_N$ type of PF spin chains \eq{b8} would be
a proper subalgebra  of the i)
$Y(gl(m))$ Yangian algebra when  $n=0$ and $m_1, \, m_2$ are positive integers, 
ii)  
$Y(gl(n))$ Yangian algebra when $m=0$ and $n_1, \, n_2$ are positive integers, 
 iii)  
$Y(gl(m|n))$ super Yangian algebra when $m, \, n$ are positive integers and 
neither the relation $m_1=n_1= 0$ nor $m_2=n_2= 0$ is satisfied. 
\begin{figure}[ht]
\vskip -1 cm 
\begin{tikzpicture}[scale=0.83]
\tikzset{
    level/.style = {
        ultra thick
    },
    connect/.style = {
        thick,
        dashed
    },
    notice/.style = {
        draw,
        rectangle callout,
        callout relative pointer={#1}
    },
    label/.style = {
        text width=2cm
    }
}

    % Draw all levels
    \draw[level] (0,0) -- node[above] {$(11|0)$} (2,0) node[right] {$10$};

    \draw[connect] (2.8,0) -- (3.5,0)  (2,0) -- (3.5,1.5) (2,0) -- (3.5,3) (2,0) -- (3.5,4.5);
    \draw[level] (3.5,0) -- node[above] {$(11|0)$} (5.5,0);
    \node[level] at (5.9,0) {$4$};
    \draw[level] (3.5,1.5) -- node[above] {$(11|1)$} (5.5,1.5);
    \node[label] at (7,1.5) {$3$};
    \draw[level] (3.5,3) -- node[above] {$(11|2),(01|0)$} (5.5,3);
    \node[label] at (7.1,3) {$2+2$};
    \draw[level] (3.5,4.5) -- node[above] {$(11|3),(01|1)$} (5.5,4.5);
    \node[label] at (7.1,4.5) {$1+4$};
    
    \draw[level] (0,3) -- node[above] {$(01|0)$} (2,3) node[right] {$8$};
    \draw[connect] (2,3) -- (3.5,3) (2,3) -- (3.5,4.5) (2,3) -- (3.5,6) ; 
    \draw[level] (3.5,6) -- node[above] {$(01|2),(10|0)$} (5.5,6);
    \node[label] at (7.1,6) {$2+2$};
    \draw[level] (3.5,7.5) -- node[above] {$(10|1)$} (5.5,7.5);
    \node[label] at (7.1,7.5) {$4$};
    
    \draw[level] (0,6) -- node[above] {$(10|0)$} (2,6) node[right] {$8$};
    \draw[connect] (2,6) -- (3.5,6) (2,6) -- (3.5,7.5) (2,6) -- (3.5,9);
    \draw[level] (3.5,9) -- node[above] {$(10|2)$} (5.5,9);
    \node[label] at (7.1,9) {$2$};
    \draw[level] (3.5,10.5) -- node[above] {$(00|1)$} (5.5,10.5);
    \node[label] at (7.1,10.5) {$1$};

    \draw[level] (0,9) -- node[above] {$(00|0)$} (2,9) node[right] {$1$};
   \draw[connect] (2,9) -- (3.5,10.5);
%    \draw[level] (3.5,12) -- node[above] {$(00|2)$} (5.5,12) node[right] {$5$};
 %   \draw[level] (3.5,13.5) -- node[above] {$(00|3)$} (5.5,13.5) node[right] {$7$};
                        
    \node[label] at (-0.5,0) {$\mc{E}=0$};
    \node[label] at (-0.5,1.5) {$\mc{E}=1$};
    \node[label] at (-0.5,3) {$\mc{E}=2$};
    \node[label] at (-0.5,4.5) {$\mc{E}=3$};
    \node[label] at (-0.5,6) {$\mc{E}=4$};
    \node[label] at (-0.5,7.5) {$\mc{E}=5$};
    \node[label] at (-0.5,9) {$\mc{E}=6$};
    \node[label] at (-0.5,10.5) {$\mc{E}=7$};

%second diagram
    \draw[level] (9.9,0) -- node[above] {$(11|0)$} (11.9,0) node[right] {$12$};
    \draw[connect] (12.6,0) -- (13.4,0)  (11.9,0) -- (13.4,1.5) (11.9,0) -- (13.4,3) (11.9,0) -- (13.4,4.5);
    \draw[level] (13.4,0) -- node[above] {$(11|0)$} (15.4,0) node[right] {$1$};
    \draw[level] (13.4,1.5) -- node[above] {$(11|1)$} (15.4,1.5) node[right] {$3$};
    \draw[level] (13.4,3) -- node[above] {$(11|2)$} (15.4,3) node[right] {$4$};
    \draw[level] (13.4,4.5) -- node[above] {$(11|3),(01|1)$} (15.4,4.5);
    \node[label] at (17,4.5) {$4+3$};
    
    \draw[level] (9.9,3) -- node[above] {$(01|0)$} (11.9,3) node[right] {$20$};
    \draw[connect] (11.9,3) -- (13.4,4.5) (11.9,3) -- (13.4,6) (11.9,3) -- (13.4,7.5); 
    \draw[level] (13.4,6) -- node[above] {$(01|2)$} (15.4,6) node[right] {$9$};
    \draw[level] (13.4,7.5) -- node[above] {$(01|3),(10|1)$} (15.4,7.5);
    \node[label] at (17,7.5) {$8+3$};
    
    \draw[level] (9.9,6) -- node[above] {$(10|0)$} (11.9,6) node[right] {$20$};
    \draw[connect] (11.9,6) -- (13.4,7.5) (11.9,6) -- (13.4,9) (11.9,6) -- (13.4,10.5);
    \draw[level] (13.4,9) -- node[above] {$(10|2)$} (15.4,9) node[right] {$9$};
    \draw[level] (13.4,10.5) -- node[above] {$(10|3)$} (15.4,10.5) node[right] {$8$};
    
    \draw[level] (9.9,9) -- node[above] {$(00|0)$} (11.9,9) node[right] {$12$};
    \draw[connect] (11.9,9) -- (13.4,12) (11.9,9) -- (13.4,13.5);
    \draw[level] (13.4,12) -- node[above] {$(00|2)$} (15.4,12) node[right] {$5$};
    \draw[level] (13.4,13.5) -- node[above] {$(00|3)$} (15.4,13.5) node[right] {$7$};
                        
    \node[label] at (9.4,0) {$\mc{E}=0$};
    \node[label] at (9.4,1.5) {$\mc{E}=1$};
    \node[label] at (9.4,3) {$\mc{E}=2$};
    \node[label] at (9.4,4.5) {$\mc{E}=3$};
    \node[label] at (9.4,6) {$\mc{E}=4$};
    \node[label] at (9.4,7.5) {$\mc{E}=5$};
    \node[label] at (9.4,9) {$\mc{E}=6$};
    \node[label] at (9.4,10.5) {$\mc{E}=7$};
    \node[label] at (9.4,12) {$\mc{E}=8$};
    \node[label] at (9.4,13.5) {$\mc{E}=9$};

    \tikzset{
    level/.style = {
        ultra thick
    },
    connect/.style = {
        thick,
        dashed
    },
    notice/.style = {
        draw,
        rectangle callout,
        callout relative pointer={#1}
    },
    label/.style = {
        text width=4cm
    }
}

    \node[label] at (2.4,-1) {$\mc{H}_3^{(3,0|0,0)}$};
    
    \node[label] at (5.9,-1) {$\mc{H}_3^{(2,1|0,0)}$};
    
    \node[label] at (12.3,-1) {$\mc{H}_3^{(2,0|2,0)}$};
    
    \node[label] at (15.8,-1) {$\mc{H}_3^{(1,1|0,2)}$};
    
\end{tikzpicture}
\caption{The spectra of some $BC_N$ type of ferromagnetic PF spin chains are 
expressed through the branched motifs and 
compared in this figure. The left diagram compares the spectra of two bosonic spin chains with Hamiltonians $\mc{H}_3^{(3,0|0,0)}$ and $\mc{H}_3^{(2,1|0,0)}$.  
Similarly, the right diagram compares the spectra of two  spin chains with Hamiltonians $\mc{H}_3^{(2,0|2,0)}$ and $\mc{H}_3^{(1,1|0,2)}$, which have both bosonic and fermionic spin degrees of freedom. Integers on the right hand side of the energy levels indicate the degeneracy factors of the corresponding branched motifs.}
\label{en_dia}
\end{figure}
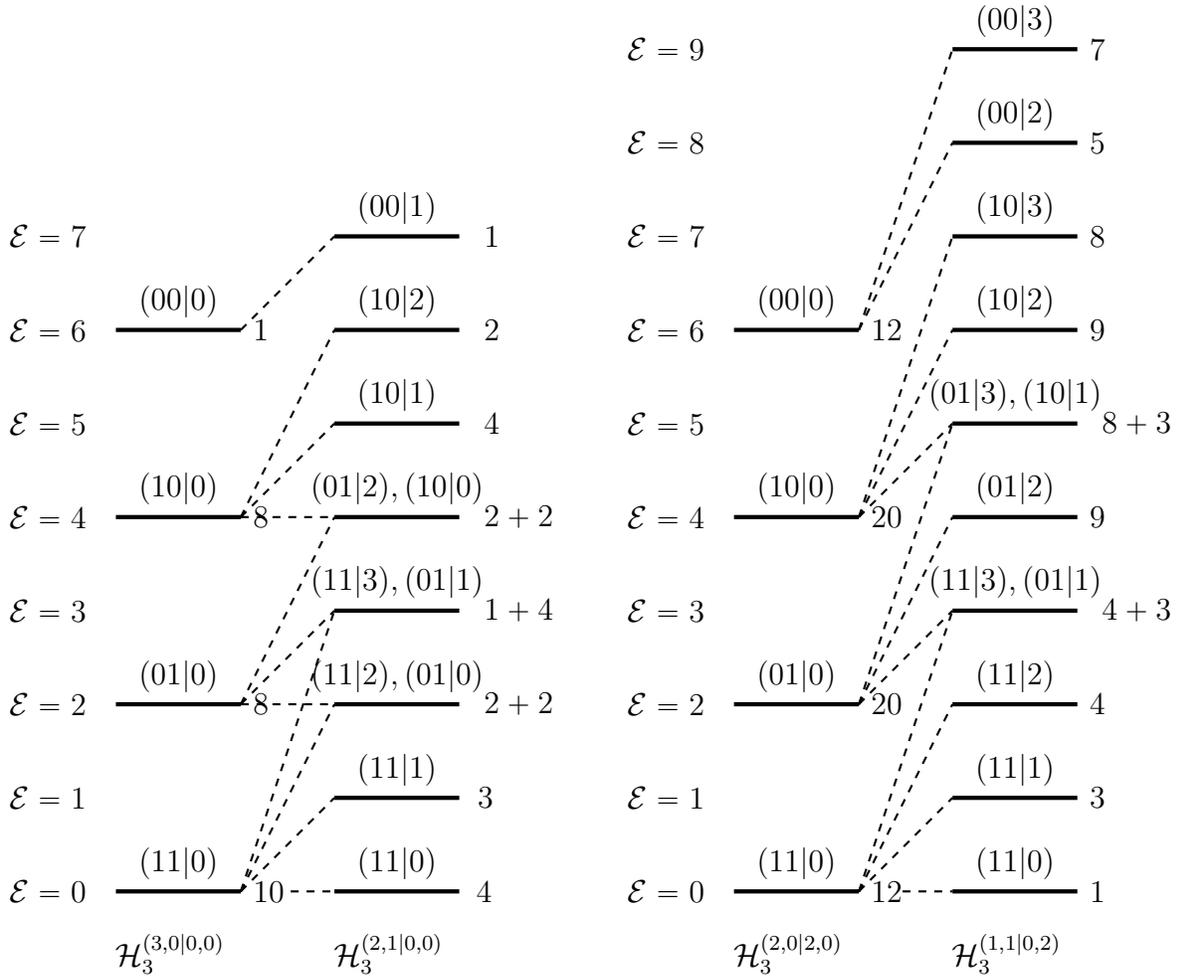

From the above discussion it is clear that,  
$BC_N$ type of PF spin chains  with    
Hamiltonians of the form   
$\mc{H}^{(m,0|n,0)}_N$
and $\mc{H}^{(0,m|0,n)}_N$ possess the maximal Yangian or super Yangian 
symmetry. Indeed, by combining Eqs.~\eq{split1} and 
\eq{eqdeg} we find the relation
\beq
 \mc{N}^{(m,0|n,0)}_{\langle k_1,k_2,....,k_r|0 \rangle} 
 = \sum\limits_{l=0}^N \mc{N}^{(m_1,m_2|n_1,n_2)}_{\langle k_1,k_2,....,k_r|l \rangle}\, ,
\label{split2}
\eeq
which, along with Eqs.~\eq{en_motif_v} and \eq{en_motif}, implies  
that each highly degenerate energy level of the Hamiltonian
$\mc{H}^{(m,0|n,0)}_N$ splits into many parts (depending on admissible 
values of $l$) and transforms into less degenerate energy levels of the 
Hamiltonian $\mc{H}^{(m_1,m_2|n_1,n_2)}_N$.
In Fig. \ref{en_dia}, we have shown how 
the spectra of some $BC_N$ type of ferromagnetic
PF spin chains with $N=3$ are 
expressed through the branched motifs and 
compared their spectra in two diagrams.
In the left diagram, we have compared the spectra of two bosonic 
spin chains with Hamiltonians  $\mc{H}_3^{(3,0|0,0)}$
and $\mc{H}_3^{(2,1|0,0)}$. In particular, we have shown that 
 the energy levels of the former Hamiltonian split due to the possible non-zero values of $l$ and transform into the energy levels of the later Hamiltonian. It may also be noted that the branched motifs $(01|3)$, $(10|3)$, $(00|0)$,
$(00|2)$ and $(00|3)$ are absent in the spectrum of the later spin chain, since the corresponding degeneracy factors are found to be zero by using \eq{Schur_wt1}. 
Similarly, in the right diagram, we have compared the spectra of two spin chains with Hamiltonians  $\mc{H}_3^{(2,0|2,0)}$
and $\mc{H}_3^{(2,0|0,2)}$ which have both bosonic and fermionic
spin degrees of freedom, and shown 
how the energy levels of the former Hamiltonian split  into the energy levels of the later Hamiltonian. The branched motifs $(01|0)$, $(10|0)$, $(00|0)$ and $(00|1)$ are absent in the spectrum of the later spin chain since the corresponding degeneracy factors are found to be zero.

 It should be noted that, in addition to the intrinsic degeneracy of the energy
 level 
  $\mc{E}^{(m_1,m_2|n_1,n_2)}_{(\de_1,\de_2,\cdots,\de_{N-1}|l)}$  
 given by $\mc{N}^{(m_1,m_2|n_1,n_2)}_ {\langle k_1, k_2,...,k_r|l\rangle}$, accidental degeneracies may also occur in the spectrum of the $BC_N$ type of PF 
 spin chain. For example, let us assume that there exist  
 two different branched motifs $(\de_1,\de_2,...,\de_{N-1}|l)$ and $(\de^{'}_1,\de^{'}_2,...,\de^{'}_{N-1}|l^{'})$
 such that 
 \[
 2\sum\limits_{j=1}^{N-1} j\de_j - l=
 2\sum\limits_{j=1}^{N-1} j\de^{'}_j - l^{'} \, .
 \]
 Since by using \eq{en_motif} one obtains
$\mc{E}^{(m_1,m_2|n_1,n_2)}_{(\de_1,\de_2,\cdots,\de_{N-1}|l)} = \mc{E}^{(m_1,m_2|n_1,n_2)}_{(\de^{'}_1,\de^{'}_2,\cdots,
\de^{'}_{N-1}|l^{'})} $,  the energy levels associated with branched motifs $(\de_1,\de_2,...,\de_{N-1}|l)$ and $(\de^{'}_1,\de^{'}_2,...,\de^{'}_{N-1}|l^{'})$ coincide with each other. In this case, the total degeneracy of the corresponding energy level becomes the sum of the degeneracy factors associated
with these two branched motifs.
In fact, for some choices of the parameters $m_1$, $m_2$, $n_1$, $n_2$
and values of $N$, there may exist multiple branched motifs with same energy level and in that case, the corresponding degeneracy will be the sum of degeneracies associated with all of these  branched motifs.
As shown in Fig.~\ref{en_dia}, due to the  accidental degeneracy in the spectrum of Hamiltonian $\mc{H}_3^{(2,1|0,0)}$, the energy levels corresponding to  the 
branched motifs $(11|2)$,  $(11|3)$ and $(01|2)$ coincide with those of the branched motifs $(01|0)$, $(01|1)$ and $(10|0)$ respectively. Similarly, 
for the case of Hamiltonian $\mc{H}_3^{(1,1|0,2)}$, the energy levels corresponding to the  
branched motifs $(11|3)$ and $(01|3)$ coincide with those of the branched motifs $(01|1)$ and $(10|1)$ respectively.  
%\vskip -2 cm

%In Figure~\ref{en_dia}, the energy levels of a $BC_N$ type of %ferromagnetic PF spin chain with $m_1=1$, $m_2=1$, $n_1=0$, $n_2=1$ %and $N=3$, characterized by the branched motifs, have been compared %to the twice of the energy levels of the corresponding $A_{N-1}$ %type of ferromagnetic PF spin chain, characterized by the original %motifs. From this diagram, it is clear that the degeneracies in the %spectrum of the $A_{N-1}$ type of ferromagnetic PF spin chain, %which are consequences of the Yangian symmetry, are reduced in the %case of its $BC_N$ version. However, in the next section, we shall %construct a one dimensional classical vertex model for this $BC_N$ %type of ferromagnetic PF spin chain.

\noi \section{\texorpdfstring{Spectra of the $BC_N$ type of anti-ferromagnetic PF spin chain}{antiferro}} \label{antiferro}
\renewcommand{\theequation}{6.{\arabic{equation}}}
\setcounter{equation}{0}
\medskip

In analogy with the ferromagnetic case, the Hamiltonian for a class of exactly solvable $BC_N$
type of anti-ferromagnetic PF spin chains with SAPSRO may be defined as
\beq
\widetilde{\mc{H}}^{(m_1,m_2|n_1,n_2)}_N
=\sum_{\underset{i \neq j}{i,j=1}}^N \left[\frac{1+ P_{ij}^{(m|n)}}{(\xi_i-\xi_j)^2} +
\frac{1+ \widetilde{{P}}_{ij}^{(m_1,m_2|n_1,n_2)}}{(\xi_i+\xi_j)^2}\right]
+\beta\sum_{i=1}^{N}\frac{1+ P_i^{(m_1,m_2|n_1,n_2)} }{\xi_i^2} \, . \label{af_hamil}
\eeq
It has been found that $BC_N$ type of SRS polynomials 
of the second kind play the role 
of generalized partition functions for these anti-ferromagnetic PF spin chains  \cite{BD17}.
In this section, at first we shall derive a recursion relation similar to \eq{RS_recursion} involving $BC_N$ type of SRS polynomials 
of the second kind. 
Subsequently, by using the above mentioned recursion relation, 
we shall express the 
later type of SRS polynomials through restricted super Schur polynomials and find out the energy levels of anti-ferromagnetic PF spin chains \eq{af_hamil}in terms of branched motifs. 

By using the freezing trick,  partition functions of the anti-ferromagnetic
PF spin chains \eq{af_hamil}  have been  derived 
in the form \cite{BD17}
\beq
\wt{\mc{Z}}^{(m_1,m_2|n_1,n_2)}_{B,N}(q)
=\sum_{\stackrel
{a_i, \, b_j, \, c_k, \,  d_l \, \in \, \Z}
{\sum\limits_{i=1}^{m_1} a_i+\sum\limits_{j=1}^{m_2} b_j
+\sum\limits_{k=1}^{n_1} c_k+\sum\limits_{l=1}^{n_2} d_l=N}}
%{a_i\geq 0,~ b_j\geq 0, ~c_k\geq 0, ~d_l\geq 0}~
\frac{\left(q^2\right)_N \cdot 
q^{\sum\limits_{i=1}^{m_1} a_i^2
+\sum\limits_{j=1}^{m_2}b_j(b_j-1)
+\sum\limits_{k=1}^{n_1} c_k}}
{\prod\limits_{i=1}^{m_1}(q^2)_{a_i}\prod\limits_{j=1}^{m_2}(q^2)_{b_j}\prod\limits_{k=1}^{n_1}(q^2)_{c_k}
\prod\limits_{l=1}^{n_2}(q^2)_{d_l}} \, .  
\label{af_part}
\eeq 
It may be noted that these partition functions can be reproduced  by taking
$x=\bar x=y=\bar y=1$ limit of the 
$BC_N$ type of SRS polynomials of the second kind given by 
\bea
&& \wt{\mbb{H}}_{B,N}^{(m_1,m_2|n_1,n_2)}
(x,\bar x;y,\bar y;q)\nn \\
&&=\sum_{\stackrel
{a_i, \, b_j, \, c_k, \,  d_l \, \in \, \Z}
{\sum\limits_{i=1}^{m_1} a_i+\sum\limits_{j=1}^{m_2} b_j
+\sum\limits_{k=1}^{n_1} c_k+\sum\limits_{l=1}^{n_2} d_l=N}} 
\hskip -1.64 cm (q^2)_N \cdot 
 q^{\sum\limits_{i=1}^{m_1}a_i^2+ \sum\limits_{j=1}^{m_2}b_j(b_j-1)
+\sum\limits_{k=1}^{n_1}c_k} \cdot 
 \prod\limits_{i=1}^{m_1}\frac{x_i^{a_i}}{(q^2)_{a_i}}
\prod\limits_{j=1}^{m_2}\frac{(\bar{x}_j)^{b_j}}{(q^2)_{b_j}}
\prod\limits_{k=1}^{n_1}\frac{y_k^{c_k}}{(q^2)_{c_k}}
\prod\limits_{l=1}^{n_2}\frac{(\bar{y}_l)^{d_l}}{(q^2)_{d_l}}
 \, .\nn \\
&&~~~~
\label{af_SRS}
\eea
Comparing \eq{SRS0} with \eq{af_SRS}, it is easy to see 
that the $BC_N$ type of SRS polynomials of the second and first kind
are related as 
\beq
\wt{\mbb{H}}_{B,N}^{(m_1,m_2|n_1,n_2)}
(x,\bar x,y,\bar y;q) = \mbb{H}_{B,N}^{(n_2,n_1|m_2,m_1)}
(\bar y,y;\bar x,x;q).
\label{f_af}
\eeq
In analogy with \eq{Gen}, the generating function $\wt{\mc{G}}_B^{(m_1,m_2|n_1,n_2)}(x,\bar{x};y,\bar{y};q;t)$ corresponding to $\wt{\mbb{H}}_{B,N}^{(m_1,m_2|n_1,n_2)}(x,\bar x;y,\bar y;q)$ may
be defined as
\beq
\wt{\mc{G}}_B^{(m_1,m_2|n_1,n_2)}(x,\bar{x};y,\bar{y};q,t)=\sum\limits_{N=0}^{\infty}\frac{
\wt{\mbb{H}}_{B,N}^{(m_1,m_2|n_1,n_2)}(x,\bar x;y,\bar y;q)}
{(q^2)_N} ~ t^N  .
\label{Gen_af}
\eeq
With the help of  Eqs.~\eq{Gen},  \eq{f_af} and \eq{Gen_af}, this generating function can be related to the generating function associated with the ferromagnetic case as 
\beq
\wt{\mc{G}}_B^{(m_1,m_2|n_1,n_2)}(x,\bar{x};y,\bar{y};q,t) = \mc{G}_B^{(n_2,n_1|m_2,m_1)}(\bar y,y;\bar x,x;q,t).
\label{f_af_gen}
\eeq
By using the above equation along with \eq{GF},  
 $\wt{\mc{G}}_B^{(m_1,m_2|n_1,n_2)}(x,\bar{x};y,\bar{y};q,t) $ 
 can be expressed as 
\beq
\wt{\mc{G}}_B^{(m_1,m_2|n_1,n_2)}(x,\bar{x};y,\bar{y};q,t)=
\mc{G}_4^{(m_1)}(x;q,t) \cdot \mc{G}_3^{(m_2)}(\bar x;q,t) \cdot 
\mc{G}_2^{(n_1)}(y;q,t) \cdot
\mc{G}_1^{(n_2)}(\bar y;q,t) \, .
\label{GF_af}
\eeq

Now, for the purpose of deriving a recursion relation for the $BC_N$ type of 
SRS polynomials of the second kind, we use 
Eqs.~\eq{g1q2}, \eq{g2q2}, \eq{g3q2}, \eq{g4q2} and \eq{GF_af}, to obtain 
a $q^2$-difference relation like 
\beq
\label{GFq2_af}
\wt{\mc{G}}_B^{(m_1,m_2|n_1,n_2)}(x,\bar x;y,\bar y;q,t)=\frac{\prod\limits_{i=1}^{m_1}(1+t~q~x_i) \prod\limits_{j=1}^{m_2}(1+t~\bar x_j)}{\prod\limits_{k=1}^{n_1}(1-t~q~y_k) \prod\limits_{l=1}^{n_2}(1-t~\bar{y}_l)}~\wt{\mc{G}}_B^{(m_1,m_2|n_1,n_2)}(x,\bar x;y,\bar y;q,q^2t).
\eeq
In analogy with \eq{factor}, we expand the prefactor in the right hand side of the above equation in a power series of $t$ as 
\beq
 \label{factor_af}
 \frac{\prod\limits_{i=1}^{m_1}(1+t~q~x_i) \prod\limits_{j=1}^{m_2}(1+t~\bar x_j)}{\prod\limits_{k=1}^{n_1}(1-t~q~y_k) \prod\limits_{l=1}^{n_2}(1-t~\bar{y}_l)} = \sum\limits_{k=0}^{\infty} ~ t^k~ \wt{E}_k^{(m_1,m_2|n_1,n_2)}(x,\bar{x};y,\bar{y};q) \, ,
\eeq
where
\beq
\label{x_qx_y_qy_af}
\wt{E}_k^{(m_1,m_2|n_1,n_2)}(x,\bar{x};y,\bar{y};q) = E_k^{(m_1+m_2|n_1+n_2)}(x,\bar{x};y,\bar{y})|_{x \rightarrow qx, y \rightarrow qy}.
\eeq
Thus $\wt{E}_k^{(m_1,m_2|n_1,n_2)}(x,\bar{x};y,\bar{y};q)$ may be considered as a $q$-deformation of the second kind of the elementary supersymmetric 
polynomial $ E_k^{(m_1+m_2|n_1+n_2)}(x,\bar{x};y,\bar{y})$. 
Expanding both sides of \Eq{GFq2_af} in powers of $t$ 
by using Eqs.~\eq{Gen_af} and \eq{factor_af},  
and subsequently comparing the powers of $t^N$, we obtain a recursion 
relation for the $BC_N$ type of SRS polynomials of the second kind as 
\bea
\label{RS_recur_af}
&& \wt{\mbb{H}}_{B,N}^{(m_1,m_2|n_1,n_2)}(x,\bar x;y,\bar y;q)\nn \\
&&= \sum\limits_{k=1}^N ~ \frac{q^{2(N-k)}\cdot(q^2)_{N-1}}{(q^2)_{N-k}}~ \wt{E}_k^{(m_1,m_2|n_1,n_2)}(x,\bar{x};y,\bar{y};q)~ 
\wt{\mbb{H}}_{B,N-k}^{(m_1,m_2|n_1,n_2)}(x,\bar x;y,\bar y;q). 
~~~~~~
\eea
The above relation can generate any $\wt{\mbb{H}}_{B,N}^{(m_1,m_2|n_1,n_2)}(x,\bar x;y,\bar y;q)$ in a recursive way from the given initial condition $\wt{\mbb{H}}_{B,0}^{(m_1,m_2|n_1,n_2)}(x,\bar x;y,\bar y;q)=1$.  %Here we have used the following notation:
%\begin{align}
%    \label{RS_noteaf}
%    \wt{\mbb{H}}_{B,N}^{(m_1,m_2|n_1,n_2)}(x,\bar x;y,\bar y;q) \equiv \wt{\mbb{H}}_{B,N}(q) && \textrm{and} && \wt{E}_k^{(m_1,m_2|n_1,n_2)}(x,\bar{x};y,\bar{y};q) \equiv \wt{E}_k(q).
%\end{align}
In analogy with the ferromagnetic case, we find that $\wt{\mbb{H}}_{B,N}^{(m_1,m_2|n_1,n_2)}(x,\bar x;y,\bar y;q)$ satisfying the recursion relation \eq{RS_recur_af} can equivalently be expressed as
\beq
\label{RS_new_af}
\wt{\mbb{H}}_{B,N}^{(m_1,m_2|n_1,n_2)}(x,\bar x;y,\bar y;q) = \sum\limits_{\Vec{k} \in \mc{P}_N} q^{ 2\sum\limits_{i=1}^{r-1} K_i} \cdot \wt{S}_{\langle k_1,...,k_r \rangle}^{(m_1,m_2|n_1,n_2)}(x,\bar x;y,\bar y;q)\, ,
\eeq
where $ \Vec{k} \equiv \{ k_1,k_2,...,k_r\}$ and $\wt{S}_{\langle k_1,...,k_r \rangle}^{(m_1,m_2|n_1,n_2)}(x,\bar x;y,\bar y;q)$ is defined as
\begin{equation}
    \wt{S}_{\langle k_1,k_2,....,k_r \rangle}^{(m_1,m_2|n_1,n_2)}(x,\bar{x};y,\bar{y};q) =
\begin{vmatrix}
\wt{E}_{k_r}(q) & \wt{E}_{k_r+k_{r-1}}(q) & \cdots & \cdots & \wt{E}_{k_r+...+k_1}(q)\\
1 & \wt{E}_{k_{r-1}}(q) & \wt{E}_{k_{r-1}+k_{r-2}}(q) & \cdots & \wt{E}_{k_{r-1}+...+k_1}(q)\\
0 & 1 & \wt{E}_{k_{r-2}}(q) & \cdots & \vdots\\
\vdots & \ddots & \ddots & \ddots & \vdots\\
0 & \cdots & 0 & 1 & \wt{E}_{k_1}(q)
\end{vmatrix} \, ,
\label{Schurder_af}
\end{equation}
where $\wt{E}_k(q) \equiv \wt{E}_k^{(m_1,m_2|n_1,n_2)}(x,\bar{x};y,\bar{y};q)  $. Using Eqs.~\eq{Schurder_af}, \eq{x_qx_y_qy_af}
and \eq{Schurdet}, 
we obtain the relation 
\beq
\label{Sxqxyqy_af}
\wt{S}_{\langle k_1,k_2,....,k_r \rangle}^{(m_1,m_2|n_1,n_2)}(x,\bar{x};y,\bar{y};q) = S_{\langle k_1,k_2,....,k_r \rangle}^{(m|n)}(x,\bar{x};y,\bar{y})|_{x \rightarrow qx, y \rightarrow qy} \, ,
\eeq
which shows that 
$\wt{S}_{\langle k_1,k_2,....,k_r \rangle}^{(m_1,m_2|n_1,n_2)}(x,\bar{x};y,
\bar{y};q)$ may be interpreted as a $q$-deformed  super Schur polynomial
of the second kind. Comparing Eq.~\eq{Sxqxyqy2} with \eq{Sxqxyqy_af}, 
we find that the $q$-deformed  super Schur polynomials
of the first kind and the second kind are related as 
\beq
\wt{S}_{\langle k_1,k_2,....,k_r \rangle}^{(m_1,m_2|n_1,n_2)}(x,\bar{x};y,\bar{y};q) 
= q^N S_{\langle k_1,k_2,....,k_r \rangle}^{(m_1,m_2|n_1,n_2)}(x,\bar{x};y,\bar{y};q^{-1}) \, .  
\eeq 
Using the above relation along with \eq{Schur_l}, one can express the 
$q$-deformed  super Schur polynomials of the second kind in terms of 
restricted super Schur polynomials as 
\beq 
\wt{S}_{\langle k_1,k_2,....,k_r \rangle}^{(m_1,m_2|n_1,n_2)}(x,\bar{x};y,\bar{y};q) 
= \sum\limits_{l=0}^N q^{N-l} \cdot S_{\langle k_1,k_2,....,k_r|l \rangle}^{(m_1,m_2|n_1,n_2)}(x,\bar{x};y,\bar{y}).   
\label{ssrss}
\eeq
Combining Eqs.~\eq{RS_new_af} and \eq{ssrss}, we finally express 
 the $BC_N$ type of  SRS polynomials of the second kind through  
restricted super Schur polynomials as 
\beq
\wt{\mbb{H}}_{B,N}^{(m_1,m_2|n_1,n_2)}(x,\bar x,y,\bar y;q) = 
\sum\limits_{\Vec{k} \in \mc{P}_N}  \sum\limits_{l=0}^N q^{2\sum\limits_{i=1}^{r-1} K_i+N-l} \cdot S^{(m_1,m_2|n_1,n_2)}_{\langle k_1,...,k_r | l \rangle}(x,\bar x,y,\bar y).
\label{RS_l_af}
\eeq
Now, putting $x= \bar x=y=\bar y=1$ into \eq{RS_l_af} and also using 
\eq{Schur_N_l}, we find that the partition functions of the $BC_N$
type of anti-ferromagnetic PF spin chains \eq{af_hamil} can be written as  
\bea
\wt{\mc{Z}}_{B,N}^{(m_1,m_2|n_1,n_2)}(q) &=& \wt{\mbb{H}}_{B,N}^{(m_1,m_2|n_1,n_2)}(x=1,\bar x=1,y=1,\bar y=1;q) \nn \\
&=& \sum\limits_{\Vec{k} \in \mc{P}_N} \sum\limits_{l=0}^N \, q^{ 2\sum\limits_{i=1}^{r-1} K_i + N - l} \cdot \mc{N}^{(m_1,m_2|n_1,n_2)}_{\langle k_1,...,k_r | l \rangle}.
\eea
It is evident that, the exponent of 
$q$ appearing in the r.h.s. of the above equation yields
the energy level  for the spin chains \eq{af_hamil}
associated with the  
branched border strip $\langle k_1,...,k_r | l \rangle$ as
\beq
\label{en_border_af}
\wt{\mc{E}}^{(m_1,m_2|n_1,n_2)}_{\langle k_1,...,k_r | l \rangle} = 2\sum\limits_{i=1}^{r-1} K_i + N - l \, ,
\eeq
and the intrinsic degeneracy of this energy level is given by $\mc{N}^{(m_1,m_2|n_1,n_2)}_{\langle k_1,...,k_r | l \rangle}$.
Moreover, using the mapping defined in \eq{bo_mot1} along with the relation 
\eq{par_motif}, the above energy level can also be written in terms of the corresponding branched motif as
\beq
\label{en_motif_af}
\wt{\mc{E}}^{(m_1,m_2|n_1,n_2)}_{(\de_1,...,\de_{N-1} | l )} = 2\sum\limits_{j=1}^{N-1} j\de_j + N - l \, . 
\eeq
%with the same degeneracy factor.

It may be noted that the intrinsic degeneracy factors corresponding to the energy levels $\mc{E}^{(m_1,m_2|n_1,n_2)}_{( \de_1,...,\de_{N-1} | l )}$ (in the ferromagnetic case) and $\wt{\mc{E}}^{(m_1,m_2|n_1,n_2)}_{( \de_1,...,\de_{N-1} | l )}$ (in the anti-ferromagnetic case) are exactly same. In addition, by summing Eqs. \eq{en_motif} and \eq{en_motif_af}, 
we obtain the relation 
\begin{equation}
\mc{E}^{(m_1,m_2|n_1,n_2)}_{( \de_1,...,\de_{N-1} | l )} + \wt{\mc{E}}^{(m_1,m_2|n_1,n_2)}_{( \de_1,...,\de_{N-1} | l )} =  2 \sum\limits_{j=1}^{N-1} j  + N = N^2 \, .
\label{energy_sum}
\end{equation}
Hence, if the spectrum of the ferromagnetic case is completely known, then one can easily obtain that of the anti-ferromagnetic case and vice-versa. Indeed,  \Eq{energy_sum} is simply a manifestation  
of the fact that the sum of two Hamiltonians, corresponding to the ferromagnetic and anti-ferromagnetic cases given by \eq{b8} and \eq{af_hamil} respectively, 
yields essentially same relation in operator form:
\begin{equation*}
\mc{H}^{(m_1,m_2|n_1,n_2)}_{N} + \wt{\mc{H}}^{(m_1,m_2|n_1,n_2)}_{N} = 2\sum\limits_{i\neq j}
\left[(\xi_i-\xi_j)^{-2}+(\xi_i+\xi_j)^{-2}\right]
+2\beta\sum\limits_i\xi_i^{-2}
=N^2,
\label{g12}
\end{equation*}
where the last sum  involving  zero points of the generalized Laguerre polynomials has been computed earlier \cite{BFGR08}. 

\noi \section{\texorpdfstring{Extended boson-fermion duality through microscopic approach}{duality}}
\renewcommand{\theequation}{7.{\arabic{equation}}}
\setcounter{equation}{0}
\medskip

The partition functions of $A_{N-1}$ type of supersymmetric 
HS and PF spin chains are known to obey boson-fermion duality relations \cite{Ha93,BUW99,HB00,BB06,BBHS07}. In the case of $A_{N-1}$ type of supersymmetric HS spin chains, the boson-fermion duality  has been derived earlier by using two different techniques: the first one uses a unitary transformation which relates the  Hamiltonians of the \su{m|n} and
\su{n|m} spin chains and corresponding partition functions, while the 
second one is a microscopic approach which utilizes a more fundamental
boson-fermion duality relation among the super Schur polynomials associated with the 
partition functions of these spin chains \cite{BBHS07}. 
In the case of $BC_N$ type of PF spin chains  \eq{b8} with SAPSRO, a   similar type of duality relation has been obtained by using the first technique as \cite{BBBD16}
\beq
\mc{Z}_{B,N}^{(m_1,m_2|n_1,n_2)}(q) = q^{N^2} \cdot \mc{Z}_{B,N}^{(n_2,n_1|m_2,m_1)}(q^{-1}) \, ,
\label{ebfd} 
\eeq
which not only involves the exchange of bosonic and fermionic degrees of freedom, but also involves the exchange of positive and negative parity degrees of freedom  associated with SAPSRO (i.e. , $m_1 \leftrightarrow m_2$ and $n_1 \leftrightarrow n_2$).
In this section, our goal is to derive such `extended' boson-fermion duality relation by using the microscopic approach. To this end, at first 
we shall derive some duality relations among the  $q$-deformed and the restricted super Schur polynomials which have been introduced in Sec.~4.
Subsequently, we shall show how the duality relation \eq{ebfd} 
emerges as a consequence of  more fundamental duality relations among the $q$-deformed and the restricted super Schur polynomials.

To begin with, let us note that the $A_{N-1}$ type of super Schur polynomials satisfy the 
boson-fermion duality relation given by \cite{BBHS07,Mo97,OAA03}  
\beq
S_{\langle \Vec{k} \rangle}^{(m|n)}(x,\bar x;y,\bar y) = S_{\langle \Vec{k}^{\, '} \rangle}^{(n|m)}(y,\bar y;x,\bar x) \, ,
\label{dual_A}
\eeq
where $\langle \Vec{k} \rangle \equiv \langle k_1,k_2,...,k_r \rangle$ is the border strip defined in Fig. \ref{bstrip} and $\langle \Vec{k}^{\, '} \rangle \equiv \langle k_1^{'},...,k_{N-r+1}^{'} \rangle$ is the corresponding conjugate border strip which is obtained by flipping $\langle \Vec{k} \rangle $
over its main diagonal. The structure of this conjugate border strip is shown in Fig.~\ref{con_dia}.
\tikzset{
    level/.style = {
        thick
    },
    connect/.style = {
        thick
    },
    notice/.style = {
        draw,
        rectangle callout,
        callout relative pointer={#1}
    },
    label/.style = {
        text width=2cm
    }
}
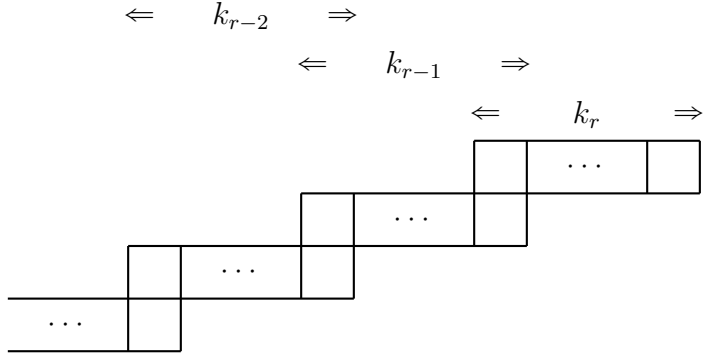
\begin{figure}
\centering
\begin{tikzpicture}
\draw[level] (0,0) -- (-3,0);
\draw[level] (0,-0.7) -- (-5.3,-0.7);
\draw[connect] (0,0) -- (0,-0.7);
\draw[connect] (-0.7,0) -- (-0.7,-0.7);
\draw[connect] (-3,0) -- (-3,-1.4);
\draw[connect] (-2.3,0) -- (-2.3,-1.4);
\node[level] at (-1.5,-0.35) {$\cdots$};
\node[level] at (-1.5,0.35) {$k_r$};
\node[level] at (-0.15,0.35) {$\Rightarrow$};
\node[level] at (-2.85,0.35) {$\Leftarrow$};

\draw[level] (-2.3,-1.4) -- (-7.6,-1.4);
\draw[connect] (-5.3,-0.7) -- (-5.3,-2.1);
\draw[connect] (-4.6,-0.7) -- (-4.6,-2.1);
\node[level] at (-3.8,-1.05) {$\cdots$};
\node[level] at (-3.8,1) {$k_{r-1}$};
\node[level] at (-2.45,1) {$\Rightarrow$};
\node[level] at (-5.15,1) {$\Leftarrow$};

\draw[level] (-4.6,-2.1) -- (-9.2,-2.1);
\draw[connect] (-7.6,-1.4) -- (-7.6,-2.8);
\draw[connect] (-6.9,-1.4) -- (-6.9,-2.8);
\node[level] at (-6.1,-1.75) {$\cdots$};
\node[level] at (-6.1,1.65) {$k_{r-2}$};
\node[level] at (-4.75,1.65) {$\Rightarrow$};
\node[level] at (-7.45,1.65) {$\Leftarrow$};

\draw[level] (-6.9,-2.8) -- (-9.2,-2.8);
\node[level] at (-8.4,-2.45) {$\cdots$};

\end{tikzpicture}
\caption{Shape of the border strip $\langle k_1^{'},k_2^{'},...,k_{N-r+1}^{'} \rangle$ conjugate to 
$\langle k_1,k_2,...,k_r \rangle$}.
\label{con_dia}
\end{figure}
Using the symmetry of  $S_{\langle \Vec{k}^{\, '} \rangle}^{(n|m)}(y,\bar y;x,\bar x)$ 
under the exchange of $x \leftrightarrow \bar{x}$ and $y \leftrightarrow \bar{y}$, \Eq{dual_A} can be also written as
\beq
S_{\langle \Vec{k} \rangle}^{(m|n)}(x,\bar x;y,\bar y) = S_{\langle \Vec{k}^{\, '} \rangle}^{(n|m)}(\bar y, y;\bar x, x) \, . 
\label{dual_A2}
\eeq
Multiplying  both sides of the above equation with $q^N$,  scaling 
the sets of variables $x$ and $y$ as $x \rightarrow q^{-1}x$ and $y \rightarrow q^{-1}y$ respectively, and finally using Eqs.~\eq{Sxqxyqy} as well as  \eq{Sxqxyqy2}, we obtain an extended boson-fermion duality relation 
between  two deformed super Schur polynomials as  
\beq
S_{\langle \Vec{k} \rangle}^{(m_1,m_2|n_1,n_2)}(x,\bar x;y,\bar y;q) = q^N \cdot S_{ \langle 
 \Vec{k}^{\, '} \rangle }^{(n_2,n_1|m_2,m_1)}(\bar y,y;\bar x,x;q^{-1}).
\eeq
Expanding both sides of the above equation by using \eq{Schur_l} and redefining the variable
$N-l$ as $l$, we obtain
%\begin{equation*}
 %\sum_{l=0}^N q^l \cdot S_{\langle k_1,...,k_r|l \rangle}
%^{(m_1,m_2|n_1,n_2)}(x,\bar x;y,\bar  y) = \sum_{l=0}^N q^{N-l} 
%\cdot S_{\langle k_1^{'},...,k_{N-r+1}^{'}|l \rangle}^{(n_2,n_1|
%m_2,m_1)}(\bar y,y;\bar x,x).
%\end{equation*}
%Redefining the variable $N-l$ as $l$ in the right hand side of the above equation, we can write
\begin{equation*}
    \sum_{l=0}^N q^l \cdot S_{\langle \Vec{k}|l \rangle}^{(m_1,m_2|n_1,n_2)}(x,\bar x;y,\bar y) = \sum_{l=0}^N q^{l} \cdot S_{\langle \Vec{k}^{\, '}|N-l \rangle}^{(n_2,n_1|m_2,m_1)}(\bar y,y;\bar x,x).
\end{equation*}
Comparing the powers of $q$ in both sides of the above equation, we find that restricted super Schur polynomials satisfy  an extended boson-fermion duality relation of the form  
\beq
\label{ebfss}
S_{\langle \Vec{k}|l \rangle}^{(m_1,m_2|n_1,n_2)}(x,\bar x;y,\bar y) = S_{\langle \Vec{k}^{\, '}|N-l \rangle}^{(n_2,n_1|m_2,m_1)}(\bar y,y;\bar x,x).
\eeq
Next, we consider the reversed border strip
$ \langle 
\vec{k}_{\mr{rev}} \rangle \equiv \langle k_r,k_{r-1},...,k_1 \rangle $
corresponding to the border strip $\langle \Vec{k} \rangle \equiv \langle k_1,k_2,...,k_r \rangle $, and  the reverse conjugate border strip 
$\langle   \Vec{k}^{\, '}_{\mr{rev}} \rangle \equiv \langle  k^{'}_{N-r+1},k^{'}_{N-r},...,k_1  \rangle$ corresponding to 
$\langle \Vec{k}^{\, '} \rangle \equiv  
\langle k_1^{'},k_2^{'},...,k^{'}_{N-r+1} \rangle$.
Following the procedure outlined in the Appendix of 
Ref.~\cite{BBCFG19a}, 
 it can be easily shown that any $q$-deformed super Schur Polynomial,  
 defined through a determinant relation of the form \eq{Schurder}, would 
 remain invariant under the reversal of the corresponding border strip:
\beq
S_{\langle \Vec{k}^{\, '} \rangle}^{(n_2,n_1|m_2,m_1)}(x,\bar{x};y,\bar{y};q) = 
S_{\langle \Vec{k}^{\, '}_{\mr{rev}}\rangle}^{(n_2,n_1|m_2,m_1)}(x,\bar{x};y,\bar{y};q) \, .
\eeq
Expanding both sides of the above equation by using \eq{Schur_l}, we find that the restricted super Schur polynomials  also remain 
invariant under the above mentioned reversal, i.e.,
\beq
S_{\langle  \Vec{k}^{\, '} |l \rangle }^{(n_2,n_1|m_2,m_1)}(x,\bar{x};y,\bar{y}) = S_{\langle \Vec{k}^{\, '}_{\mr{rev}}
  |l \rangle }^{(n_2,n_1|m_2,m_1)}(x,\bar{x};y,\bar{y}) \, .
\label{r_schur_sym}
\eeq
Combining Eqs.~\eq{ebfss} and \eq{r_schur_sym}, we 
obtain an extended boson-fermion duality relation of the form  
\beq
\label{ebfss1}
S_{\langle \Vec{k}|l \rangle}^{(m_1,m_2|n_1,n_2)}(x,\bar x;y,\bar y) = S_{\langle \Vec{k}^{\, '}_{\mr{rev}}|N-l \rangle}^{(n_2,n_1|m_2,m_1)}(\bar y,y;\bar x,x).
\eeq

%Substituting the above equation into the SRS polynomial given by %\eq{RS_l} and redefining the variable $N-l$ as $l$, we obtain
%\beq
%\mbb{H}_{B,N}^{(m_1,m_2|n_1,n_2)}(x,\bar x,y,\bar y;q) = \sum
%\limits_{\Vec{k} \in \mc{P}_N} \sum\limits_{l=0}^N q^{N^2 - 
%2\sum\limits_{i=1}^{r-1} K_i - l} \cdot S^{(n_2,n_1|m_2,m_1)}
%_{\langle \Vec{k}^{\, '} | l \rangle}(\bar y,y;\bar x,x).
%\label{SRS_red}
%\eeq

For our purpose of proving 
\Eq{ebfd}, it is also needed to find out a relation between the partial sums 
associated with the border strip
$\langle \Vec{k} \rangle$  and those of the corresponding 
reversed conjugate border strip
$\langle \Vec{k}^{\, '}_{\mr{rev}} \rangle$. To this end
it may be noted that, the set of partial sums $\{ K_1^{'},K_2^{'},...,K^{'}_{N-r+1} \}$ corresponding to the conjugate border strip $\langle \Vec{k}^{\, '} \rangle$ are given by \cite{BBHS07}
\beq
\{ K_1^{'},K_2^{'},...,K^{'}_{N-r} \} = \{ N-K_{r+1},N-K_{r+2},...,N-K_N \}\, ,
\label{con_par}
\eeq
where $\{ K_{r+1},K_{r+2},...,K_N \}$ represents the set of complementary partial sums corresponding to the border strip $\langle \Vec{k} \rangle$, i.e.,
\beq
\{ K_{r+1},K_{r+2},...,K_N \} = \{ 1,2,...,N \} - \{ K_{1},K_{2},...,K_{r} \}.
\label{comp_par}
\eeq
Using Eqs.~\eq{con_par} and \eq{comp_par}, and inserting
$K_r=N$, we get
\beq
\sum_{i=1}^{N-r} K_i^{'}  = N(N-r) - \sum_{i=r+1}^{N} K_i 
= N(N-r) - \frac{N(N-1)}{2} + \sum_{i=1}^{r-1} K_i.
\label{par_conju}
\eeq
Next, we note that 
the sums of the partial sums for the border strip $\langle \Vec{k} \rangle$ and the corresponding reversed border strip $ \langle 
\vec{k}_{\mr{rev}} \rangle$
%$ \langle 
%\vec{k}_{\mr{rev}} \rangle \equiv \langle k_r,k_{r-1},...,k_1 \rangle $
 can be written as
\begin{align}
    \sum\limits_{i=1}^{r-1} K_i & = k_1 + (k_1+k_2) + \cdots + (k_1+
    \cdots +k_{r-1}) 
    %= rk_1+(r-1)k_2+\cdots+k_r 
    = \sum\limits_{j=1}^{r} (r-j)k_j \, ,
    \label{border_sum}
\end{align}
and
\begin{align}
    \sum\limits_{i=1}^{r-1} \wt{K}_i & = k_r + (k_r+k_{r-1}) + \cdots + (k_r
    + \cdots +k_2) 
    %= rk_r+(r-1)k_{r-1}+\cdots+k_1 
    = \sum\limits_{j=1}^{r} (j-1)k_j \, ,
    \label{reverse_sum}
\end{align}
respectively, 
where $\wt{K}_i$ denotes  the $i$-th partial sum associated with the 
reversed border strip  $ \langle \vec{k}_{\mr{rev}} \rangle$.
Adding Eqs.~\eq{border_sum} and \eq{reverse_sum}, we obtain    
the relation 
\beq
\sum\limits_{i=1}^{r-1} K_i + \sum\limits_{i=1}^{r-1} \wt{K}_i = N(r-1)\, .
\eeq
It is evident that, for the case of conjugate border strip 
$\langle \Vec{k}^{\, '} \rangle$ and the reversed conjugate border strip 
$\langle   \Vec{k}^{\, '}_{\mr{rev}} \rangle$,  
the above relation can be written as  
\beq
\sum\limits_{i=1}^{N-r} K^{'}_i +
\sum\limits_{i=1}^{N-r} \wt{K}_i^{\hskip .007 cm '} = N(N-r)\, ,
\label{conj_rev}
\eeq
where $\wt{K}_i^{\hskip .007 cm '}$ denotes  the $i$-th partial sum associated with the  reversed conjugate border strip 
$\langle   \Vec{k}^{\, '}_{\mr{rev}} \rangle$.
Combining \eq{conj_rev} and \eq{par_conju}, we finally obtain
a relation between the partial sums 
associated with  border strip
$\langle \Vec{k} \rangle$  and those of 
$\langle \Vec{k}^{\, '}_{\mr{rev}} \rangle$ as
\beq
\sum_{i=1}^{r-1} K_i = \frac{N(N-1)}{2} - 
\sum_{i=1}^{N-r} \wt{K}_i^{\hskip .007 cm '} \,  .
\label{par_rev_sum}
\eeq

Substituting the relations \eq{ebfss1} and \eq{par_rev_sum} into 
\Eq{RS_l}, and making a change of the summation variables, we obtain
\bea
 &&\mbb{H}_{B,N}^{(m_1,m_2|n_1,n_2)}(x,\bar x;y,\bar y;q) \nn \\
 &&~~~~= q^{N^2} \cdot \sum\limits_{\Vec{k}^{\, '}_{\mr{rev}}  \in \mc{P}_N} \sum\limits_{l=0}^N 
 \left(q^{-1}\right)^{\big\{N(N-1) - 2\sum\limits_{i=1}^{N-r} \wt{K}_i^{\hskip .007 cm '} + l \, \big\} }\cdot S^{(n_2,n_1|m_2,m_1)}_{\langle \Vec{k}^{\, '}_{\mr{rev}} |
 l \rangle}(\bar y,y;\bar x,x).
\label{SRS_dual}
\eea
It is interesting to observe that, the power of $q$ in \Eq{RS_l} and the power of 
$q^{-1}$ in \Eq{SRS_dual} can be represented through  
essentially same function of the variables
$ \Vec{k} $  and 
$ \Vec{k}^{\, '}_{\mr{rev}} $ respectively.
Consequently, \Eq{SRS_dual} can be expressed in the form
\beq
\mbb{H}_{B,N}^{(m_1,m_2|n_1,n_2)}(x,\bar x;y,\bar y;q) = q^{N^2} \cdot \mbb{H}_{B,N}^{(n_2,n_1|m_2,m_1)}(\bar y,y;\bar x,x;q^{-1}).
\eeq
Inserting $x=\bar{x}=y =\bar{y}=1$ to the above equation  and using \eq{SRSpa},  we finally obtain the extended boson-fermion duality relation \eq{ebfd}.
Using the results of section \ref{antiferro}, it can be shown that the partition functions associated with  $BC_N$ type of 
anti-ferromagnetic PF spin chains obey similar type of duality relation. 

\noi \section{\texorpdfstring{Concluding remarks}{Conclusion}}
\renewcommand{\theequation}{8.{\arabic{equation}}}
\setcounter{equation}{0}
\medskip
Here we establish that the  spectra of the $BC_N$ type of PF spin chains with SAPSRO, including the degeneracy factors associated with all energy levels,  can be described completely through the branched motifs. Recently introduced $BC_N$ type of multivariate SRS polynomials, which are closely connected
with the partition functions of the above mentioned spin chains, play the central role in our approach. At first, we show that these SRS polynomials satisfy the  recursion relation \eq{RS_recursion} involving
 a particular type of $q$-deformation of the elementary supersymmetric   polynomials. With the help
 of such $q$-deformed elementary supersymmetric  polynomials, subsequently we define a few  mathematical entities which would be helpful for further analysis. For example,
we use a Jacobi-Trudi like formula  \eq{Schurder} to define the corresponding $q$-deformed super Schur polynomials. Moreover, by expanding these $q$-deformed super Schur polynomials in powers of $q$, we obtain the so called restricted super Schur polynomials. With the help of skew Young tableaux
 associated with border strips, we also present some combinatorial expressions \eq{comb_q} and \eq{Schur_wt} for the $q$-deformed and the restricted super Schur polynomials respectively. 
 
 Applying the  recursion relation \eq{RS_recursion}, 
 subsequently we derive novel expressions \eq{RS_new} and  \eq{RS_l}
 for  the $BC_N$ type of SRS polynomials through suitable linear combinations
 of the $q$-deformed  and the 
 restricted super Schur polynomials respectively. As shown in \Eq{en_motif},
 the later expressions for the $BC_N$ type of SRS polynomials enable us to describe the spectra of the corresponding ferromagnetic PF spin chains with $N$ number of lattice sites in terms of the branched motifs like 
$(\de_1, \de_2,...,\de_{N-1}|l)$, where $\de_i \in \{ 0,1 \}$ and $ l \in \{ 0,1,...,N \}$. 
By taking  a particular limit of the  restricted super Schur polynomials, we 
obtain the combinatorial expression  \eq{Schur_wt1} which determines the intrinsic degeneracy of the energy level 
%$\mc{E}^{(m_1,m_2|n_1,n_2)}_{(\de_1, \de_2,...,\de_{N-1}|l)}$
associated with the branched motif
$(\de_1, \de_2,...,\de_{N-1}|l)$. 
However, as discussed in Section 5, there may exist more than one 
branched motifs with coincident energy levels. In that case,
 the total degeneracy of the coincident energy level is determined by 
the sum of the intrinsic 
degeneracy factors corresponding to those branched motifs. 
%$\mc{N}^{(m_1,m_2|n_1,n_2)}_ {(\de_1, \de_2,...,\de_{N-1}|l)} 
%\equiv\mc{N}^{(m_1,m_2|n_1,n_2)}_{\langle k_1,...,k_r | l \rangle},$
 In analogy with the ferromagnetic case, we also obtain a complete 
 classification for  the spectra of 
 the $BC_N$ type of anti-ferromagnetic PF spin chains through the 
 branched motifs. Moreover, we derive an extended boson-fermion
 duality relation \eq{ebfss}
 among the restricted super Schur polynomials and show that the 
 partition functions of the $BC_N$ type of PF spin chains exhibit a 
 similar duality relation.

 From  \Eq{Schur_wt1} one can easily see that the intrinsic 
degeneracy of the energy level associated with the branched motif 
$(\de_1, \de_2,...,\de_{N-1}|l)$ depends  
on the discrete parameters $m_1, \, m_2, \, n_1, \, n_2$ of the 
Hamiltonian $\mc{H}^{(m_1,m_2|n_1,n_2)}_N$  in  \eq{b8}. 
 %If this degeneracy factor becomes zero for some choice of the discrete
% parameters, then the corresponding energy level would be forbidden  from %the spectrum of the spin chain.  
It may also be noted that, 
 \Eq{split1}  establishes an interesting connection between  
the intrinsic degeneracy factors corresponding to the energy levels of the 
$BC_N$ type Hamiltonian  
$\mc{H}^{(m_1,m_2|n_1,n_2)}_N$ and those of the 
%$Y(gl(m|n))$ super Yangian invariant
$A_{N-1}$ type of  Hamiltonian $\mc{H}_N^{(m|n)}$   in 
\eq{a1}, where $m=m_1+m_2$ and $n=n_1+n_2$. 
By using the above mentioned equations, we find that  
the spectrum of $\mc{H}^{(m,0|n,0)}_N$ in \eq{ys}
obeys a selection rule which 
forbids the occurrence  of branched motifs like  $(\de_1, \de_2,...,\de_{N-1}|l)$for $l > 0$. Moreover, the nontrivial intrinsic degeneracy of the branched motif  $(\de_1, \de_2,...,\de_{N-1}|0)$ coincides with that of the usual motif 
 $(\de_1, \de_2,...,\de_{N-1})$ associated with the Hamiltonian $\mc{H}_N^{(m|n)}$. Since the latter Hamiltonian exhibits the $Y(gl(m|n))$  Yangian 
 symmetry, our analysis indirectly proves that the Hamiltonian
 $\mc{H}^{(m,0|n,0)}_N$ (and also the related Hamiltonian
 $\mc{H}^{(0,m|0,n)}_N$) has the same symmetry.  
 Our analysis using Eqs.~\eq{Schur_wt1} and  \eq{split1}  
 also suggests that, except for the above mentioned special cases
 where either $m_1=n_1=0$ or $m_2=n_2=0$, 
 the symmetry algebra of the  Hamiltonian  
$\mc{H}^{(m_1,m_2|n_1,n_2)}_N$ would be a proper subalgebra of the 
$Y(gl(m|n))$  Yangian algebra.
Even though the symmetry algebras of
some spin Calogero models associated 
with the $BC_N$ root system have been studied earlier \cite{Ya95,VN08},  
as far as we know the symmetry algebras of the $BC_N$ type of 
PF spin chains like \eq{b8}
have not received much attention. So it might be interesting 
as a future study to find out the symmetry algebras of these 
$BC_N$ type of spin chains 
and explore how the representations of such symmetry algebras are  
connected with the $q$-deformed and the restricted super Schur polynomials.

Finally, we would like to make a few comments about some further   
developments which we are making at present by using the results of this paper. First of all, it is possible to extend the connection between  
$BC_N$ type of multivariate SRS polynomials and the partition functions 
of the $BC_N$ type of PF spin chains even in the presence of the chemical 
potentials. By using such connection, one can express the complete spectra 
and partition functions of the $BC_N$ type of PF spin chains with chemical potentials in terms of some one-dimensional classical vertex models.
Moreover, by employing the transfer matrices associated with those classical vertex models, one can study various thermodynamic properties and criticality of the $BC_N$ type of PF spin chains with chemical potentials. We plan to describe these interesting developments about vertex models and thermodynamic properties of these spin chains in some forthcoming publications.

\bigskip 
\noi {\bf Acknowledgements}
\medskip

One of the authors (BBM) would like to thank Artemio Gonz\'alez-L\'opez and Federico Finkel for fruitful discussions and kind hospitality at Universidad Complutense de Madrid during ARTEMP'2019 Conference. 
%This work is supported by the TPAES 
%project of Theory Division, 
%Saha Institute of Nuclear Physics, India.

%\newpage 

\bibliographystyle{model1a-num-names}
\bibliography{cmprefs}

\end{document}